\pdfoutput=1
\documentclass[prl,aps,superscriptaddress,footinbib,twocolumn]{revtex4-2}
\usepackage{color}
\usepackage{amsmath}
\usepackage{amsthm}
\usepackage{amssymb}
\usepackage{stmaryrd}
\usepackage{graphicx}
\usepackage{mdframed}
\usepackage{paracol}
\usepackage{siunitx}
\usepackage{comment}
\usepackage[export]{adjustbox}
\usepackage[unicode=true,
 bookmarks=true,bookmarksnumbered=false,bookmarksopen=false,
 breaklinks=false,pdfborder={0 0 1},backref=false,colorlinks=true]
 {hyperref}
\hypersetup{
 linkcolor=magenta, urlcolor=blue, citecolor=blue, pdfstartview={FitH}, hyperfootnotes=false, unicode=true}

\makeatletter

\usepackage{amsfonts}\usepackage{tabularx}\usepackage{dcolumn}\usepackage{bm}\usepackage{graphicx}\usepackage{epstopdf}\usepackage{tikz}
\usepackage{times}
\usepackage{cleveref}
\hypersetup{urlcolor=blue}

\makeatother

\newmdtheoremenv{theo}{Theorem}

\begin{document}


\title{Quantum-classical gravity distinction in reservoir-engineered massive quantum system}

\author{Ziqian Tang}\thanks{These authors contributed equally to this work}
\affiliation{Beijing Key Laboratory of Fault-Tolerant Quantum Computing, Beijing Academy of Quantum Information Sciences, Beijing 100193, China}
\affiliation{Bejing National Laboratory for Condensed Matter Physics, Institute of Physics, Chinese Academy of Sciences, Beijing 100190, China}
\affiliation{University of Chinese Academy of Sciences, Beijing 100049, China}

\author{Zizhao Han}\thanks{These authors contributed equally to this work}
\affiliation{Center for Quantum Information, IIIS, Tsinghua University, Beijing 100084, China}

\author{Zikuan Kan}
\affiliation{School of Physics, Renmin University of China, Beijing 100872, China}

\author{Chen Yang}
\affiliation{School of Physics, Peking University, Beijing 100871, China}

\author{Zeji Li}
\affiliation{School of Integrated Circuits, Tsinghua University, Beijing 100084, China}

\author{Yining Jiang}
\affiliation{Beijing Key Laboratory of Fault-Tolerant Quantum Computing, Beijing Academy of Quantum Information Sciences, Beijing 100193, China}

\author{Yulong Liu}
\email{liuyl@baqis.ac.cn}
\affiliation{Beijing Key Laboratory of Fault-Tolerant Quantum Computing, Beijing Academy of Quantum Information Sciences, Beijing 100193, China}

\date{\today}

\begin{abstract}
Massive quantum systems have emerged as compelling tabletop interface-systems for testing the quantum nature of gravity. However, conventional schemes that focus on directly using gravity to induce entanglement suffer from overwhelming environmental decoherence: maintaining entanglement between two oscillators requires an impractically high mechanical quality factor. In this work, we put forward an alternative reservoir-engineered scheme, whose core function is to quantify how gravity modifies (rather than prepares) the steady-state entanglement. Compared to quantum gravity, classical gravity introduces additional dissipative channels, which in turn give rise to distinct entanglement characteristics and thus enable the discrimination between the two types of gravity. Notably, this entanglement difference can still be maintained even when the mechanical quality factor is far below the threshold required by conventional schemes. Moreover, it demonstrates significant robustness against non-gravitational couplings, specifically, those like Casimir and Coulomb forces that are inherent in experimental setups. Our scheme relaxes the experimental requirements for verifying quantum gravity, thereby paving a new path toward its near-term realization.

\end{abstract}

\maketitle


Understanding whether gravity is fundamentally quantum remains one of the most profound open questions in modern physics. As experimental techniques improve, tabletop tests involving massive quantum systems have emerged as promising candidates to explore physics at the quantum–gravity interface \cite{santos2017optomechanical,li2018measurements,westphal2021measurement,whittle2021approaching,liu2021gravitational,youssefi2023squeezed,fuchs2024measuring,bose2025massive,marletto2025quantum,carney2019tabletop,gallerati2022interaction,tang2025cavity}. A particularly compelling hypothesis suggests that if gravity is quantum, then it should be able to mediate entanglement  on its own between two spatially separated mechanical oscillators \cite{Feynman1957,PhysRevLett.119.240401,PhysRevLett.119.240402}. However, detecting such entanglement directly is severely hindered by environmental decoherence. When $k_B T \gg \hbar \omega_m$, a universal constraint $2 \gamma_m k_B T \lesssim \hbar G \Lambda \rho$ must be satisfied to preserve gravity-induced entanglement against thermal decoherence from the environment, where $G$ is the Newton's constant, $\omega_m$ and $\gamma_m$ are the oscillators' mechanical frequency and damping rate, $T$ is the environmental temperature, $\rho$ is the mass density, and $\Lambda$ is a geometry-dependent form factor \cite{ miao2020quantum,tang2025optimal}. At environmental temperature around 10~\si{mK}, achieving and maintaining gravitationally mediated entanglement for spherical oscillators requires mechanical dissipation $\gamma_m /2\pi \lesssim 10^{-16}~\si{Hz}$, which means an impractically high mechanical quality factor—a daunting challenge beyond current experimental reach \cite{krisnanda2020observable,plato2023enhanced,kotler2021direct,margalit2021realization}.  In addition, non-gravitational interactions between the oscillators, such as Casimir forces, electrostatic Coulomb forces, or dipole–dipole forces, can mimic or obscure the gravitational signal \cite{Casimir:1948dh,buhmann2013dispersion,klimchitskaya2009casimir,van2020quantum,wang2021strong}. Eliminating these spurious effects typically demands stringent control over oscillator geometry, size, materials, and shielding, further compounding the experimental complexity. Subject to this two issues, experimental research in this direction has been severely limited.

In this work, we propose an experimental scheme based on a reservoir-engineered massive quantum system to circumvent the aforementioned difficulties. Instead of observing entanglement generated solely by gravity, we utilize an engineered reservoir to first establish steady-state entanglement between two massive oscillators, and then investigate how this entanglement is modified by gravitational interactions. Notably, quantum and classical gravity induce qualitatively distinct steady-state entanglement shifts: quantum gravity preserves the entanglement, while classical gravity introduces an effective dissipative channel to degrade it \cite{diosi1987universal,penrose1996gravity,ghirardi1986unified,bassi2023collapse,yang2013macroscopic,helou2017measurable,scully2022semiclassical,liu2023semiclassical,liu2025semiclassical,kafri2014classical,kafri2014bounds,aleksandrov1981statistical,carlip2008quantum,oppenheim2023postquantum,miki2025role,kryhin2025distinguishable}. This distinction provides a clear signature for discriminating between the two scenarios. The scheme has two key advantages. First, since gravity only needs to perturb pre-established entanglement, constraints on mechanical quality factor are significantly relaxed. Second, non-gravitational couplings are naturally suppressed by the oscillators’ frequency difference, allowing us to neglect them even when their strength far exceeds that of gravity. These advancements substantially boost the feasibility of near-term quantum gravity tests using massive optomechanical systems.

\textit{Gravity models}---In a quantum framework, gravity acts as a coherent interaction channel that can, in principle, generate entanglement. By contrast, if a gravity model is categorized as purely classical, it cannot, on its own, generate entanglement between two subsystems. Such interactions are operationally equivalent to local operations with classical communication (LOCC)~\cite{nielsen2010quantum,chitambar2014everything}. A representative model that matches the LOCC condition is the one introduced by Kafri, Taylor, and Milburn (KTM model) \cite{kafri2014classical}. They treat gravity as a continuous measurement process followed by classical feedback. The model reproduce Newtonian forces on average, and inevitably introduce decoherence that will weaken existed quantum correlations. In our analysis, we adopt the KTM framework as a classical benchmark. While not unique, it captures the essential structure of LOCC classical gravity models. For more general LOCC models, they share the feature of introducing additional dissipation~\cite{angeli2025probing}, which shifts entanglement differently from quantum models and thereby creates conditions for our scheme to be effective in their distinction.  

\begin{figure}[t]
    \centering    
    \hspace*{-0.2cm} 
    \vspace*{-0.2cm}
    \includegraphics[width=1\linewidth]{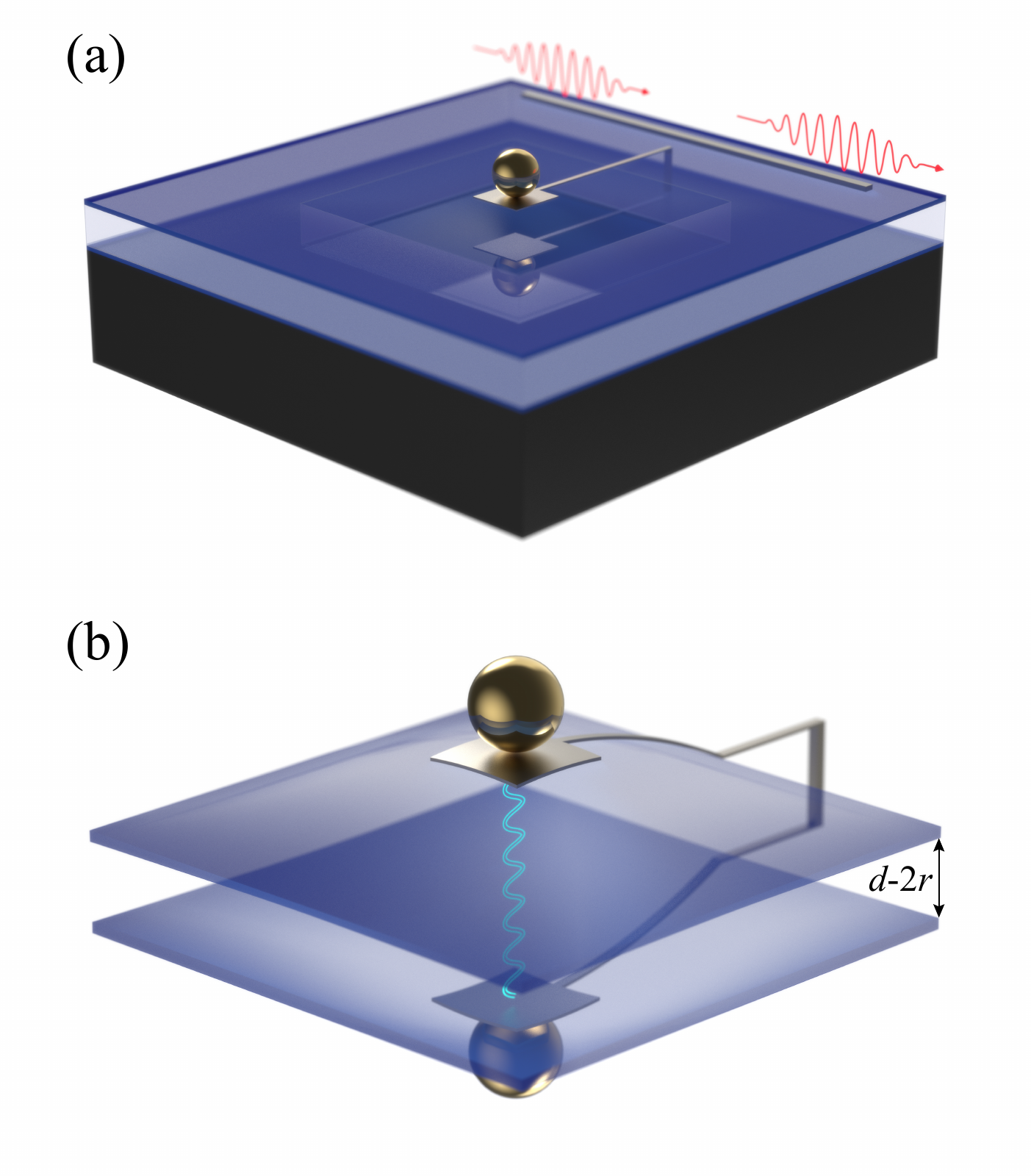}
    \caption{The proposed experiment setup. (a) A pair of spherical gold masses is connected to the plates of a capacitor and loaded onto membrane oscillators. The plates act as part of a microwave cavity formed by an on-chip superconducting circuit, mediating coupling between the mechanical oscillation and the cavity modes. The red wave represents the pump beam, which excites the mode in the microwave cavity through electromagnetic induction. (b) A schematic of the oscillator when its oscillation displaces it from equilibrium. The blue wavy line denotes the gravitational interaction between the masses. At equilibrium, the center-to-center distance of the two spheres is $d$, each sphere has radius $r$, and the separation of the membranes is $d – 2r$.
} 
    \label{fig:device}
\end{figure} 

To briefly describe the KTM model and lay the groundwork for our subsequent approach, we consider a two-object system, where each object can be treated as a point mass. These objects reach equilibrium under mutual gravitational interaction and other forces, and undergo small oscillations about their equilibrium positions. In the KTM model, the position of each oscillator is continuously measured by a hypothetical observer; the measurement outcomes are then used to control the motion of the other oscillator, thereby reproducing the dynamics of classical Newtonian gravity within the LOCC framework. From this model, we derive that the density matrix $\hat{\rho}$ of the system evolves in accordance with the master equation given by:

\begin{equation}
    \label{eq:KTM_mastereq}
    \begin{split}
    \frac{d\hat{\rho}}{dt}=&-\frac{i}{\hbar}[\hat{H}_0,\hat{\rho}]-\frac{i}{\hbar}K_G[\hat{x}_1\hat{x}_2,\hat{\rho}]-\sum_{k=1}^2\frac{\Gamma_k}{2\hbar}[\hat{x}_k,[\hat{x}_k,\hat{\rho}]]\\
    &-\frac{K_G^2}{8\hbar\Gamma_1}[\hat{x}_2,[\hat{x}_2,\hat{\rho}]]-\frac{K_G^2}{8\hbar\Gamma_2}[\hat{x}_1,[\hat{x}_1,\hat{\rho}]].\\
    \end{split}
\end{equation}
Here, $K_G = 2G m_1 m_2 /d^3$ represents the magnitude of the gravitational gradient, where $m_1$ and $m_2$ are the masses of the two objects, and $d$ is their equilibrium separation. $\Gamma_k$ ($k = 1, 2$) are model parameters, characterize the rate at which information is gained through hypothetical measurements. In Eq.~\eqref{eq:KTM_mastereq}, the first two terms correspond to the conventional quantum mechanical Hamiltonian evolution, while the remaining three terms represent additional dissipation arising from classical gravity. More specifically, the third term represents measurement-induced decoherence, which increases with the measurement rate $\Gamma_k$; the last two terms correspond to feedback noise caused by measurement inaccuracy, which decreases as the measurement rate increases. Mathematical analysis shows that the additional dissipation induced by classical gravity is minimized when $\Gamma_1 = \Gamma_2 = K_G/2$; yet it remains nonzero, thereby leading to a finite gap relative to the purely quantum case. This gap gives rise to observable effects, which in turn yield a discrimination signature for distinguishing between classical and quantum gravity models.

\textit{Reservoir-engineered massive quantum system in the presence of gravity}---We propose an on-chip integrated quantum-gravity interface based on mass-loaded membrane mechanical oscillators. Specifically, we consider a system of two coaxially aligned thin-film membranes [schematically shown in Fig.~\ref{fig:device}(a)], each loaded with a massive sphere of mass $M$, which form two mechanical oscillators with natural frequencies $\omega_a$ and $\omega_b$. Owing to their mutual gravitational interaction and other interactions, their frequencies shift to $\omega'_{a (b)} = \sqrt{\omega_{a (b)}^2 - K_M / M}$. Here, $K_M = K_G + K_{\text{others}}$, where $K_G = \partial^2 V_G / \partial x_1 \partial x_2|_0$ and $K_{\text{others}} = \partial^2 V_{\text{others}}/ \partial x_1 \partial x_2|_0$ denote the cross-gradient of the gravitational and non-gravitational interaction potentials $V_G$ and $V_{\text{others}}$, respectively evaluated at their equilibrium positions.

A microwave cavity with frequency $\omega_c$ is formed by a superconducting on-chip circuit [as shown in Fig.~\ref{fig:device}(b)] \cite{aspelmeyer2014cavity,seis2022ground,liu2023coherent,liu2025degeneracy}. The positions of the oscillators modulate the capacitance of the circuit, thereby inducing optomechanical coupling. The single-photon coupling strengths between the cavity mode and the two oscillators are $g_a$ and $g_b$, respectively. The corresponding Hamiltonian is given by: 
\begin{equation}
    \begin{split}
        \hat{H} &= \hbar \omega'_{a} \hat{a}^\dagger \hat{a} + \hbar \omega'_{b} \hat{b}^\dagger \hat{b} + \hbar g_a (\hat{a}^\dagger + \hat{a})\hat{c}^\dagger \hat{c} + \hbar g_b (\hat{b}^\dagger + \hat{b})\hat{c}^\dagger \hat{c}\\
        &+ \hat{H}_M.
    \end{split}
\end{equation}
where $\hat{a}$ and $\hat{b}$ are the annihilation operators for the two mechanical modes, and $\hat{c}$ is for the cavity mode. $\hat{H}_M = \hbar k_M (\hat{a}^\dagger + \hat{a})(\hat{b}^\dagger + \hat{b})$ describes the interaction between the two masses with $k_M = K_M x_{\text{zpf},a}x_{\text{zpf},b} /\hbar$. Here, $x_{\text{zpf},a(b)} = \sqrt{\hbar / 2 M {\omega'}_{a(b)}}$ are the zero point fluctuations of the oscillators $a$ and $b$. The system dynamics are then governed by the following master equation (see Supplemental Material S1 for details): 
\begin{equation}
    \begin{split}
        \dot{\hat{\rho}} &= -\frac{i}{\hbar}[\hat{H}, \hat{\rho}] + \gamma_a (\overline{n}_a + 1) \mathcal{D}[\hat{a}]\hat{\rho} + \gamma_a \overline{n}_a \mathcal{D}[\hat{a}^\dagger]\hat{\rho}\\
        &+ \gamma_b (\overline{n}_b + 1) \mathcal{D}[\hat{b}]\hat{\rho} + \gamma_b \overline{n}_b \mathcal{D}[\hat{b}^\dagger]\hat{\rho} + \kappa \mathcal{D}[\hat{c}]\hat{\rho}\\
        &+2\kappa_{G,a}\mathcal{D}[\hat{a}^\dagger + \hat{a}]\hat{\rho}+2\kappa_{G,b}\mathcal{D}[\hat{b}^\dagger + \hat{b}]\hat{\rho}.
    \end{split}
\end{equation}
Here, \(\mathcal{D}[\hat{O}]\hat{\rho}=\hat{O}\hat{\rho}\hat{O}^\dagger-\frac12\bigl\{\hat{O}^\dagger\hat{O},\hat{\rho}\bigr\}\) is the Lindblad super-operator. $\gamma_{a(b)}$ and $\overline{n}_{a(b)}$ represent the mechanical dissipation rate and phonon occupation number of the oscillator $a$ and $b$, $\kappa$ represents the dissipation rate of the cavity while $\kappa_{G,a}$ and $\kappa_{G,b}$ are the dissipation rates of the gravity, which equal to zero for quantum gravity and equal to $\kappa_{G,a(b)} = [\Gamma_{a(b)} /\hbar + K_G^2/4\hbar\Gamma_{b(a)}]x_{\text{zpf},a(b)}^2$ for classical gravity in KTM model. With the symmetry assumption $\Gamma_a=\Gamma_b$, they have lower bounds $\kappa_{G,a(b)} \geq K_G x_{\text{zpf},a(b)}^2/\hbar$.

When two pump beams with frequency $\omega_c\pm\omega_m$ are injected, they introduce an additional term $\hat{H}_\text{drive} = \hbar (\mathcal{E}_+^* e^{i\omega_m t} + \mathcal{E}_-^* e^{-i\omega_m t})e^{i\omega_c t}\hat{c} + \text{h.c.}$ to the Hamiltonian, and generate a cavity field, $\omega_m = (\omega'_a + \omega'_b)/2$ is the central frequency of the mechanical modes and $\mathcal{E}_{\pm}$ are the pump amplitudes. Following the standard results of two-mode squeezing reservoir-engineered optomechanics \cite{woolley2014two}, we work in the rotating frame defined by $\hat{U} = e^{i \hat{H}_0 t /\hbar}$, where $\hat{H}_0 = \hbar \omega_c \hat{c}^\dagger \hat{c} + \hbar \omega_m \hat{a}^\dagger \hat{a} + \hbar \omega_m \hat{b}^\dagger \hat{b}$. Under the strong driving approximation $\mathcal{E}_\pm \gg \omega_m$, resolved sideband approximation $\kappa \ll \omega_{a(b)}$, and adiabatic limit $\kappa > \Omega, G_\pm$, where $\Omega = (\omega'_a - \omega'_b)/2$, $G_\pm = (g_a + g_b)\overline{c}_\pm /2$ denote the many-photon optomechanical couplings, and $\overline{c}_\pm = i\mathcal{E}_\pm / (\pm i\omega_m - \kappa/2)$ the steady-state amplitudes of the fields at the driven sidebands, the effective master equation is obtained as follows:
\begin{equation}
    \begin{split}
        \dot{\hat{\rho}} &= -\frac{i}{\hbar}[\hbar \Omega (\hat{a}^\dagger \hat{a} - \hat{b}^\dagger \hat{b}) + \hbar k_M (\hat{a}^\dagger \hat{b} + \hat{a}\hat{b}^\dagger), \hat{\rho}]\\
        &+ \gamma_a (\overline{n}_a + 1) \mathcal{D}[\hat{a}]\hat{\rho} + \gamma_a \overline{n}_a \mathcal{D}[\hat{a}^\dagger]\hat{\rho}\\
        &+ \gamma_b (\overline{n}_b + 1) \mathcal{D}[\hat{b}]\hat{\rho} + \gamma_b \overline{n}_b \mathcal{D}[\hat{b}^\dagger]\hat{\rho}\\
        &+ \Gamma \mathcal{D}[\hat{\beta}_1 + \hat{\beta}_2]\hat{\rho}\\
        &+ 2(\kappa_{G,a}\mathcal{D}[\hat{a}]\hat{\rho}+ \kappa_{G,a}\mathcal{D}[\hat{a}^\dagger]\hat{\rho} + \kappa_{G,b}\mathcal{D}[\hat{b}]\hat{\rho}+ \kappa_{G,b}\mathcal{D}[\hat{b}^\dagger]\hat{\rho}).
    \end{split}
\end{equation}
Here $\hat{\beta}_1 = \hat{a}\cosh r + \hat{b}^\dag\sinh r$ and $\hat{\beta}_2 = \hat{b}\cosh r + \hat{a}^\dag\sinh r$ are the mechanical two-mode Bogoliubov operators, with the parameter $r = \tanh^{-1} (G_+ / G_-)$. Additionally, $\Gamma = 4\mathcal{G}^2 / \kappa$ denotes the optomechanical damping rate, where $\mathcal{G} = \sqrt{G_-^2 - G_+^2}$ is an effective optomechanical coupling. We take the single-photon coupling strengths to be equal, i.e., $g_a=g_b=g$.

Assuming $\gamma_a = \gamma_b = \gamma_m$, $\overline{n}_a = \overline{n}_b = \overline{n}$, and $\kappa_{G,a} = \kappa_{G,b} = \kappa_G$, and the mechanical dissipation is low enough that $\gamma_m \ll \Omega$, one obtains the steady-state logarithmic negativity \cite{horodecki2009quantum} for the subsystem $a$ and $b$ as follows:
\begin{equation}
    \label{eq:EN}
    \begin{split}
        E_\mathcal{N}& \approx \max \Bigg\{0,\  -\ln \Bigg\{\frac{(2\overline{n}+1)\gamma_m +4\kappa _G+\Gamma  e^{-2 r}}{\gamma_m +\Gamma }\\
        &+\frac{\Gamma}{4 (\gamma_m +\Gamma )^3 (\xi+\Gamma  \cosh 2r)} \Bigg[\Gamma\left(4 \xi^2+\gamma_m ^2+\Gamma ^2\right)\\
        & (\xi+\gamma_m )^2 (\xi-\Gamma ) \tanh r-(\xi-\gamma_m )^2 (\xi+\Gamma) \coth r\\
        &+ 2\xi (\eta \sinh 2r -2\gamma_m\Gamma e^{-2r} + 2\Gamma^2 \cosh 2r )\\
        &+\Gamma(\Gamma ^2\cosh 4r -\eta e^{-4r})\Bigg]\Bigg(\frac{k_M}{\Omega}\Bigg)^2 \Bigg\}\Bigg\},
    \end{split}
\end{equation}
where $\xi = (2\overline{n} +1)\gamma_m + 4\kappa_G$ and $\eta= \gamma_m ^2+4 \gamma_m  \Gamma +2\Gamma ^2$. Details of the derivation are provided in the Supplemental Material S2. Using Eq.~\eqref{eq:EN}, the entanglement difference between the quantum and classical gravity models can be obtained by subtracting the case of $\kappa_G = 0$ (quantum gravity) from that of $\kappa_G \neq 0$ (classical gravity).


\textit{Plausible experimental parameters and simulation results}---We assume the mechanical oscillators $a$ and $b$ have resonance frequencies $\omega_{a,b}/2\pi = 50 \pm 0.5~\si{Hz}$, and the cavity dissipation rate is $\kappa/2\pi = 10~\si{Hz}$, such that the system operates in the resolved-sideband regime. The environment is maintained at a temperature of $T = 10~\si{mK}$, corresponding to a thermal phonon occupation number $\overline{n} = 1/[\exp(\hbar \omega_m / k_B T) - 1] \approx 4 \times 10^6$. We consider the oscillators to be made of gold, which has density $\rho = 19.3~\si{g/cm^3}$. We assume that the spherical oscillators are very close to each other, so that the form factor can be approximated as $\Lambda\approx\pi/3$~\cite{krisnanda2020observable,miao2020quantum}, yielding a gravitational coupling rate of $k_G/2\pi \approx \kappa_G/2\pi \approx 7 \times 10^{-10}~\si{Hz}$. Here, $k_G = K_G x_{\text{zpf},a}x_{\text{zpf},b} /\hbar$. We have summarized all parameters in Supplemental Material S3.

\begin{figure}[ht]
    \centering 
    \vspace*{0.24cm}
    \hspace*{-0.8cm} 
    \vspace*{0.1cm}
    \includegraphics[width=1\linewidth]{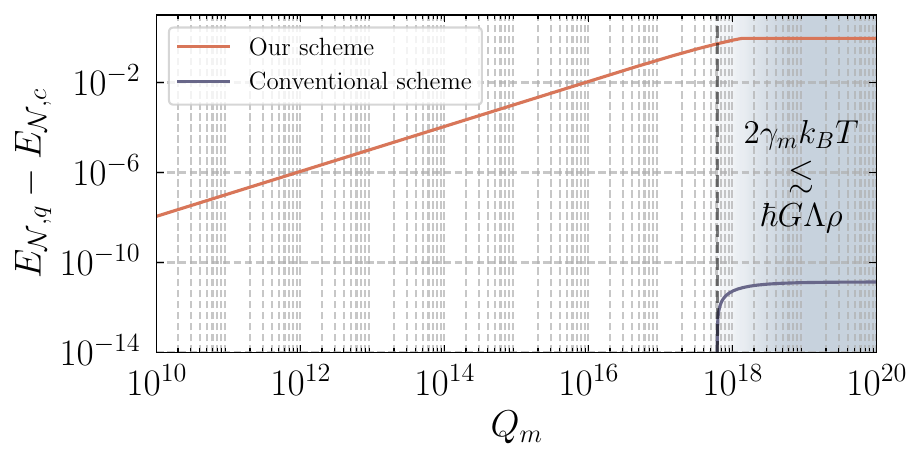}
    \caption{Simulated entanglement differences between quantum and classical gravity models under our scheme and the conventional one, respectively, plotted against mechanical quality factor $Q_m$. The gray thick vertical dashed line marks the threshold for the existence of entanglement in the conventional scheme. We choose the parameters such that, while the mechanical quality $Q_m$ equals $2.5\times 10^{10}$, the single-photon optomechanical couplings $g/(2\pi)= 1~\si{Hz}$ and the pump beam amplitude $|\mathcal{E}_+| = 10^2~\mathrm{Hz}$, $|\mathcal{E}_-| = 2\times 10^2~\mathrm{Hz}$. The phases of $\mathcal{E}_{\pm}$ are set such that \( \overline{c}_{\pm} \) are real numbers. The pump amplitude will be changed while $Q_m$ increases, due to the parameter adjusting rule.
} 
    \vspace{-0.30cm}
    \label{fig:gap}
\end{figure}

Fig.~\ref{fig:gap} represents the difference in entanglement between the quantum gravity model and the classical gravity model under our scheme and under the conventional scheme, in which the two oscillators are expected to be entangled directly via gravity \cite{krisnanda2020observable}. In our simulations, we treat the mechanical quality factor $Q_m=\omega_m/\gamma_m$ as a variable, with $\omega_m$ fixed and $\gamma_m$ adjusted. We adjust $\Gamma$ concurrently such that the ratio $\gamma_m/\Gamma$ remains constant. This can be accomplished by decreasing the pump amplitudes $\mathcal{E}_{\pm}$ as $\gamma_m$ is reduced.  

Fig.~\ref{fig:gap} shows that the entanglement difference in our scheme is always substantially larger than in the conventional scheme. Notably, in the conventional scheme, once the quality factor $Q_m$ falls below a certain threshold, no entanglement can be generated, making it impossible to distinguish between quantum and classical gravity models. In contrast, our scheme maintains the ability to discriminate between the two models over a much wider parameter range, thereby greatly relaxing the experimental requirements for such discrimination, particularly the constraint on the mechanical quality factor. For example, under our parameters, a mechanical quality factor of  $Q_m \sim 10^{10}$ would achieve a difference in entanglement surpassing conventional schemes — a level that is nearly attainable in practice~\cite{engelsen2024ultrahigh}. The simulation in Fig.~\ref{fig:gap} also reveals that for $Q_m<10^{18}$ , the entanglement difference in our scheme is approximately proportional to $Q_m$. This behavior can also be derived by directly computing the difference in Eq.~\eqref{eq:EN} between the $\kappa_G = 0$ and $\kappa_G \neq 0$ cases under certain approximations:
\begin{equation}
    \label{eq:gap}
    \begin{split}
        E_{\mathcal{N},q}-E_{\mathcal{N},c}\approx \frac{4\kappa_G}{(2\overline{n}+1)\gamma_m + \Gamma e^{-2r}} \propto \frac{1}{\gamma_m} \propto Q_m
    \end{split}
\end{equation}
Details of this approximate formula are provided in the Supplemental Material S4. 

Furthermore, we observe that the entanglement difference saturates (to approximately 1.0) as \(Q_m\) exceeds $10^{18}$. This saturation arises because our parameter-tuning rule causes \(\Gamma\) to decrease in proportion to \(\gamma_m\). Consequently, according to Eq.~\eqref{eq:EN}, the relative weight of the additional dissipation \(\kappa_G\) in the classical model increases, eventually leading the classical model’s entanglement to zero. Once this regime is reached, further increasing $Q_m$ (reducing \(\gamma_m\)) no longer inducing a significant change in the entanglement difference.

\begin{figure}[t]
    \centering   
    \vspace*{0.6mm}
    \hspace*{-0.8cm} 
    \vspace*{0.5mm}
    \includegraphics[width=0.98\linewidth]{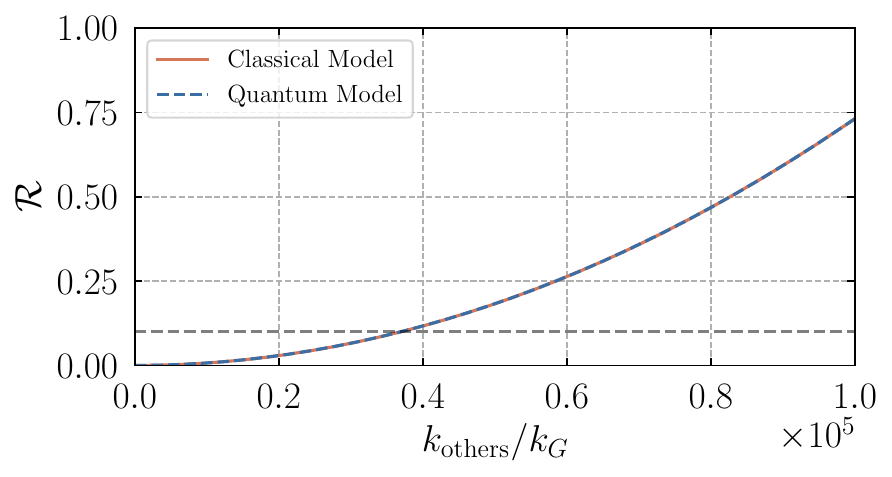}
    \caption{Ratio $\mathcal{R}$ of the entanglement contributed by non-gravitational coupling to the entanglement difference between quantum and classical gravity models, plotted against the non-gravitational coupling strength (normalized by the gravitational coupling strength). The numerator of the ratio was simulated under both quantum and classical gravity models, but the difference between them is negligible. The gray thick horizontal dashed line marks the $0.1$ level of the ratio, below which the non-gravitational coupling can be regarded as insignificant. Here, $Q_m$ is set to $2.5\times 10^{10}$, consistent with the reference value shown in Fig.~\ref{fig:gap}.
} 
    \vspace*{-1.4mm}
    \label{fig:non_grav}
\end{figure}

We finally assess the robustness of our scheme against non-gravitational inter-oscillator couplings (beyond the desired gravitational interaction). According to Eq.~\eqref{eq:EN}, the influence of the coupling strength on entanglement appears as the term $(k_M/\Omega)^2$, which is naturally suppressed as the inter-oscillator frequency difference $\Omega$ increases. We define $\mathcal{R}=(E_{\mathcal{N}}-E_{\mathcal{N},k_{\text{others}}=0})/(E_{\mathcal{N},q}-E_{\mathcal{N},c})$ as the ratio of the entanglement shift induced by non-gravitational couplings to the entanglement difference between the two models. As shown in Fig.~\ref{fig:non_grav}, the ratio $\mathcal{R}$ remains small even when those non-gravitational couplings are much stronger than the gravitational coupling. This result can also be rationalized via an order-of-magnitude analysis of Eq.~\eqref{eq:EN}. Under our choice of parameters, the magnitude of the coefficient of $(k_M/\Omega)^2$ term is approximately 0.8, thus the contributions to the expression inside the logarithm from classical gravitational dissipation and the inter-oscillator coupling scale respectively as $\sim 4\kappa_G/[(2\overline{n}+1)\gamma_m + \Gamma e^{-2r}]$ (as shown in Eq.~\eqref{eq:gap}) and $\sim 0.8(k_M/\Omega)^2$. When non-gravitational couplings are non-negligible, these two contributions become comparable. Using $\kappa_G \approx k_G$ and $k_{\text{others}} = k_M - k_G$, we estimate that $k_{\text{others}}/k_G \sim \Omega/\sqrt{0.2k_G[(2\overline{n}+1)\gamma_m + \Gamma e^{-2r}]} \sim 10^5$, which is consistent with the values in our simulations where the influence of $k_{\text{others}}$ becomes non-negligible. This result demonstrates that our scheme is highly robust against non-gravitational couplings, significantly alleviating the stringent ``coupling-cleanliness" requirements typically imposed on experimental tests of quantum gravity.


\textit{Conclusion}---In this work, we propose an on-chip integrated quantum-gravity interface based on mass-loaded membrane mechanical oscillators. Utilizing this setup allows us to first prepare the massive mechanical oscillators into an entangled state through thermal reservoir engineering. We then investigate the gravitational interactions between the oscillators under both quantum and classical gravity models. Our comparative analysis reveals that quantum and classical models of gravity influence steady-state entanglement in qualitatively distinct ways: while quantum gravity can preserve steady-state entanglement, classical gravity introduces an effective dissipative channel that diminishes it. This distinction provides a clear signature to differentiate between the two gravitational frameworks.

Compared to conventional proposals that directly utilize gravitational interactions to induce entanglement between mechanical resonators, our approach significantly relaxes the constraints on mechanical quality factor, as gravitational interactions only need to perturb pre-established entanglement rather than independently counteract environmental decoherence. Additionally, non-gravitational couplings are inherently suppressed by the inter-oscillator frequency difference, allowing us to disregard their influence even when their strengths exceed those of gravitational interactions.

Ultimately, our proposal addresses one of the most profound questions in modern physics: whether gravity is fundamentally quantum. By easing the experimental requirements for verifying quantum gravity, this work paves the way for its near-term realization in tabletop experiments.

\textit{Acknowledgment}---We thank Prof. Haixing Miao for fruitful discussions.  We also thank R. Yang for preparing Figure~\ref{fig:device}. Y. Liu acknowledges the support of Beijing Natural Science Foundation (Z240007), National Natural Science Foundation of China (No.~92365210 and No.~12374325), Young Elite Scientists Sponsorship Program by CAST (Grant No.~2023QNRC001), and Beijing Municipal Science and Technology Commission (Grant No.~Z221100002722011). Z. Han is supported by the National Natural Science Foundation of China (Grants No.~T2225008), the Innovation Program for Quantum Science and Technology (No.~2021ZD0302203), Tsinghua University Dushi Program, and the Shanghai Qi Zhi Institute Innovation Program SQZ202318.


\bibliography{apssamp}

\end{document}


\setcounter{secnumdepth}{3}
\renewcommand{\thesection}{S\arabic{section}}
\renewcommand{\thesubsection}{S\arabic{section}.\arabic{subsection}}
\renewcommand{\thesubsubsection}{S\arabic{section}.\arabic{subsection}.\arabic{subsubsection}}

\makeatletter
\def\@seccntformat#1{\csname the#1\endcsname\quad}
\makeatother


\title{Supplemental Material for ``Quantum-classical gravity distinction in reservoir-engineered massive quantum system"}

\author{Ziqian Tang}\thanks{These authors contributed equally to this work}
\affiliation{Beijing Key Laboratory of Fault-Tolerant Quantum Computing, Beijing Academy of Quantum Information Sciences, Beijing 100193, China}
\affiliation{Bejing National Laboratory for Condensed Matter Physics, Institute of Physics, Chinese Academy of Sciences, Beijing 100190, China}
\affiliation{University of Chinese Academy of Sciences, Beijing 100049, China}

\author{Zizhao Han}\thanks{These authors contributed equally to this work}
\affiliation{Center for Quantum Information, IIIS, Tsinghua University, Beijing 100084, China}

\author{Zikuan Kan}
\affiliation{School of Physics, Renmin University of China, Beijing 100872, China}

\author{Chen Yang}
\affiliation{School of Physics, Peking University, Beijing 100871, China}

\author{Zeji Li}
\affiliation{School of Integrated Circuits, Tsinghua University, Beijing 100084, China}

\author{Yining Jiang}
\affiliation{Beijing Key Laboratory of Fault-Tolerant Quantum Computing, Beijing Academy of Quantum Information Sciences, Beijing 100193, China}

\author{Yulong Liu}
\email{liuyl@baqis.ac.cn}
\affiliation{Beijing Key Laboratory of Fault-Tolerant Quantum Computing, Beijing Academy of Quantum Information Sciences, Beijing 100193, China}

\date{\today}

\maketitle

\section{Derivation of the effective master equation}

In this section, we derive the crucial dynamical equation of the main text--the effective two-mode master equation. We start from the original three-mode master equation~\cite{campaioli2024quantum}:

\begin{equation}
    \begin{split}
        \dot{\hat{\rho}} &= -\frac{i}{\hbar}[\hat{H}, \hat{\rho}] + \gamma_a (\overline{n}_a + 1) \mathcal{D}[\hat{a}]\hat{\rho} + \gamma_a \overline{n}_a \mathcal{D}[\hat{a}^\dagger]\hat{\rho}+ \gamma_b (\overline{n}_b + 1) \mathcal{D}[\hat{b}]\hat{\rho} + \gamma_b \overline{n}_b \mathcal{D}[\hat{b}^\dagger]\hat{\rho} + \kappa \mathcal{D}[\hat{c}]\hat{\rho}\\
        &+2\kappa_{G,a}\mathcal{D}[\hat{a}^\dagger + \hat{a}]\hat{\rho}+2\kappa_{G,b}\mathcal{D}[\hat{b}^\dagger + \hat{b}]\hat{\rho}.
    \end{split}
\end{equation}
Here

\begin{equation}
\begin{split}
\hat{H} &= \hbar \omega'_{a} \hat{a}^\dagger \hat{a} + \hbar \omega'_{b} \hat{b}^\dagger \hat{b} + \hbar g_a (\hat{a}^\dagger + \hat{a})\hat{c}^\dagger \hat{c} + \hbar g_b (\hat{b}^\dagger + \hat{b})\hat{c}^\dagger \hat{c}+ \hat{H}_\text{drive} + \hat{H}_M
\end{split}
\end{equation}
with $\hat{H}_M = \hbar k_M (\hat{a}^\dagger + \hat{a})(\hat{b}^\dagger + \hat{b})$ and $\hat{H}_\text{drive} = \hbar (\mathcal{E}_+^\dagger e^{i\omega_m t} + \mathcal{E}_-^\dagger e^{-i\omega_m t})e^{i\omega_c t}\hat{c} + \text{h.c.}$. The physical meaning of all terms and parameters is given in the main text.

In the rotating frame defined by $\hat{H}_0 = \hbar \omega_m \hat{a}^\dagger \hat{a} + \hbar\omega_m \hat{b}^\dagger \hat{b} + \hbar\omega_c \hat{c}^\dagger \hat{c}$ with $\omega_m = (\omega'_{a} + \omega'_{b})/2$, the system’s master equation can be written as
\begin{equation}
    \begin{split}
    \label{eq:me0}
        \dot{\hat{\rho}} &= -\frac{i}{\hbar}[\hat{H}', \hat{\rho}] + \gamma_a (\overline{n}_a + 1) \mathcal{D}[\hat{a}]\hat{\rho} + \gamma_a \overline{n}_a \mathcal{D}[\hat{a}^\dagger]\hat{\rho} + \gamma_b (\overline{n}_b + 1) \mathcal{D}[\hat{b}]\hat{\rho} + \gamma_b \overline{n}_b \mathcal{D}[\hat{b}^\dagger]\hat{\rho}\\
        &+ \kappa \mathcal{D}[\hat{c}]\hat{\rho} +2\kappa_{N,a}\mathcal{D}[\hat{a}^\dagger e^{i\omega_m t} + \hat{a} e^{-i\omega_m t}]\hat{\rho}+2\kappa_{N,b}\mathcal{D}[\hat{b}^\dagger e^{i\omega_m t} + \hat{b} e^{-i\omega_m t}]\hat{\rho}\\
    \end{split}
\end{equation}
where
\begin{equation}
    \begin{split}
    \hat{H}' &= \hbar \Omega (\hat{a}^\dagger \hat{a} - \hat{b}^\dagger \hat{b}) + \hbar g_a (\hat{a}e^{-i\omega_m t} + \hat{a}^\dagger e^{i\omega_m t})\hat{c}^\dagger \hat{c}+ \hbar g_b (\hat{b}e^{-i\omega_m t} + \hat{b}^\dagger e^{i\omega_m t})\hat{c}^\dagger \hat{c} + \hbar (\mathcal{E}_+^* e^{i\omega_m t}  + \mathcal{E}_-^* e^{-i\omega_m t})\hat{c}\\
    &+ \hbar (\mathcal{E}_+ e^{-i\omega_m t} + \mathcal{E}_- e^{i\omega_m t})\hat{c}^\dagger + \hbar k_M (\hat{a}^\dagger e^{i\omega_m t} + \hat{a} e^{-i\omega_m t})(\hat{b}^\dagger e^{i\omega_m t} + \hat{b} e^{-i\omega_m t})
    \end{split}
\end{equation}
and $\Omega = (\omega'_{a} - \omega'_{b})/2$. 

For the dissipative terms in Eq.~\eqref{eq:me0} arising from the classical gravity model, we can expand it and, under the rotating-wave approximation, neglect the high-frequency terms to obtain $ \mathcal{D}[\hat{a}^\dagger e^{i\omega_m t} + \hat{a} e^{-i\omega_m t}]\hat{\rho}\approx \mathcal{D}[\hat{a}]\hat{\rho}+ \mathcal{D}[\hat{a}^\dag]\hat{\rho}$ and $ \mathcal{D}[\hat{b}^\dagger e^{i\omega_m t} + \hat{b} e^{-i\omega_m t}]\hat{\rho}\approx \mathcal{D}[\hat{b}]\hat{\rho}+ \mathcal{D}[\hat{b}^\dag]\hat{\rho}$. This leads to the master equation: 

\begin{equation}
    \begin{split}
    \label{eq:me}
        \dot{\hat{\rho}} &= -\frac{i}{\hbar}[\hat{H}', \hat{\rho}] + \gamma_a (\overline{n}_a + 1) \mathcal{D}[\hat{a}]\hat{\rho} + \gamma_a \overline{n}_a \mathcal{D}[\hat{a}^\dagger]\hat{\rho} + \gamma_b (\overline{n}_b + 1) \mathcal{D}[\hat{b}]\hat{\rho} + \gamma_b \overline{n}_b \mathcal{D}[\hat{b}^\dagger]\hat{\rho}\\
        &+ \kappa \mathcal{D}[\hat{c}]\hat{\rho} +2\kappa_{N,a}(\mathcal{D}[\hat{a}]\hat{\rho}+ \mathcal{D}[\hat{a}^\dag]\hat{\rho})+2\kappa_{N,b}(\mathcal{D}[\hat{b}]\hat{\rho}+ \mathcal{D}[\hat{b}^\dag]\hat{\rho}).
    \end{split}
\end{equation}

The Heisenberg Langevin equations of the system, neglecting noise terms, is then given by \cite{breuer2002theory}:

\begin{equation}
\begin{split}
\label{eq:HL}
    &\dot{\hat{a}}=-i\Omega\hat{a}-i g_a e^{i\omega_mt}\hat{c}^\dag\hat{c}-i k_M(\hat{b}+\hat{b}^\dag e^{2i\omega_mt})-\frac{\gamma_a}{2}\hat{a},\\
    &\dot{\hat{b}}=i\Omega\hat{b}-i g_b e^{i\omega_mt}\hat{c}^\dag\hat{c}-i k_M(\hat{a}+\hat{a}^\dag e^{2i\omega_mt})-\frac{\gamma_b}{2}\hat{b},\\
    &\dot{\hat{c}}=-i\left[g_a(\hat{a}e^{-i\omega_mt}+\hat{a}^\dag e^{i\omega_mt})+g_b(\hat{b}e^{-i\omega_mt}+\hat{b}^\dag e^{i\omega_mt})\right]\hat{c}-i(\mathcal{E}_+e^{-i\omega_mt}+\mathcal{E}_-e^{i\omega_mt})-\frac{\kappa}{2}\hat{c}.
\end{split}
\end{equation}

Assuming resolved sidebands ($\omega_{a(b)}'\gg \kappa$), we separate the high-frequency components of $\hat{c}$ \cite{woolley2008nanomechanical}:

\begin{equation}
    \hat{c}(t)=\hat{c}_0(t)+\hat{c}_+(t)e^{-i\omega_mt}+\hat{c}_-(t)e^{i\omega_mt}.
\end{equation}
Substituting into Eq.~\eqref{eq:HL} yields the new set of equations:

\begin{equation}
    \begin{split}
    \label{eq:HL2}
        &\dot{\hat{a}}=-i\Omega\hat{a}-i g_a e^{i\omega_mt}(\hat{c}_0^\dag+\hat{c}_+^\dag e^{i\omega_mt}+\hat{c}_-^\dag e^{-i\omega_mt})(\hat{c}_0+\hat{c}_+e^{-i\omega_mt}+\hat{c}_-e^{i\omega_mt})-i k_M(\hat{b}+\hat{b}^\dag e^{2i\omega_mt})-\frac{\gamma_a}{2}\hat{a},\\
        &\dot{\hat{b}}=i\Omega\hat{b}-i g_b e^{i\omega_mt}(\hat{c}_0^\dag+\hat{c}_+^\dag e^{i\omega_mt}+\hat{c}_-^\dag e^{-i\omega_mt})(\hat{c}_0+\hat{c}_+e^{-i\omega_mt}+\hat{c}_-e^{i\omega_mt})-i k_M(\hat{a}+\hat{a}^\dag e^{2i\omega_mt})-\frac{\gamma_b}{2}\hat{b},\\ 
        &\dot{\hat{c}}_0+\dot{\hat{c}}_+e^{-i\omega_mt}-i\omega_m\hat{c}_+e^{-i\omega_mt}+\dot{\hat{c}}_-e^{i\omega_mt}+i\omega_m\hat{c}_ie^{i\omega_mt}\\
        =&
        -i\left[g_a(\hat{a}e^{-i\omega_mt}+\hat{a}^\dag e^{i\omega_mt})+g_b(\hat{b}e^{-i\omega_mt}+\hat{b}^\dag e^{i\omega_mt})\right](\hat{c}_0+\hat{c}_+e^{-i\omega_mt}+\hat{c}_-e^{i\omega_mt})-i\left(\mathcal{E}_+e^{-i\omega_mt}+\mathcal{E}_-e^{i\omega_mt}\right)\\
        &-\frac{\kappa}{2}\left(\hat{c}_0+\hat{c}_+e^{-i\omega_mt}+\hat{c}_-e^{i\omega_mt}\right).
    \end{split}
\end{equation}
In the first two equations of Eq.~\eqref{eq:HL2}, we average out the high-frequency terms. For the third equation, it can be separated into three equations corresponding to $\dot{\hat{c}}_0$, $\dot{\hat{c}}_+$, and $\dot{\hat{c}}_-$, each containing only the terms in the original equation that are frequency-matched. The resulted equations are

\begin{equation}
    \begin{split}
    \label{eq:HL3}
        &\dot{\hat{a}}=-i\Omega\hat{a}-i g_a(\hat{c}_0^\dag \hat{c}_++\hat{c}_-^\dag \hat{c}_0)-i k_M \hat{b}-\frac{\gamma_a}{2}\hat{a},\\
        &\dot{\hat{b}}=i\Omega\hat{b}-i g_b(\hat{c}_0^\dag \hat{c}_++\hat{c}_-^\dag \hat{c}_0)-i k_M \hat{a}-\frac{\gamma_b}{2}\hat{b},\\
        &\dot{\hat{c}}_0=-i\hat{c}_-(g_a\hat{a}+g_b\hat{b})-i\hat{c}_+(g_a\hat{a}^\dag+g_b\hat{b}^\dag)-\frac{\kappa}{2}\hat{c}_0,\\
        &\dot{\hat{c}}_+=i\omega_m\hat{c}_+-i\hat{c}_0(g_a \hat{a}+g_b \hat{b})-i\mathcal{E}_+-\frac{\kappa}{2}\hat{c}_+,\\
        &\dot{\hat{c}}_-=-i\omega_m\hat{c}_--i\hat{c}_0(g_a^\dag \hat{a}+g_b^\dag \hat{b})-i\mathcal{E}_--\frac{\kappa}{2}\hat{c}_-.\\
    \end{split}
\end{equation}
In the last two equations of Eq.~\eqref{eq:HL3}, using the strong-pump condition, the terms proportional to $g_{a(b)}$ can be neglected compared with the $\mathcal{E}_{+(-)}$ terms, so the steady state of $\hat{c}_{+(-)}$ can be solved directly: $\bar{c}_+=\frac{i\mathcal{E}_+}{i\omega_m-\frac{\kappa}{2}}$, $\bar{c}_-=\frac{i\mathcal{E}_-}{-i\omega_m-\frac{\kappa}{2}}$. Substituting the steady value of $\bar{c}_{+(-)}$ into the first three equations, we obtain:

\begin{equation}
    \begin{split}
    \label{eq:HL4}
         &\dot{\hat{a}}=-i\Omega\hat{a}-i g_a(\bar{c}_+\hat{c}_0^\dag+\bar{c}_-^* \hat{c}_0)-i k_M \hat{b}-\frac{\gamma_a}{2}\hat{a},\\
         &\dot{\hat{b}}=i\Omega\hat{b}-i g_b(\bar{c}_+\hat{c}_0^\dag +\bar{c}_-^* \hat{c}_0)-i k_M \hat{a}-\frac{\gamma_b}{2}\hat{b},\\
         &\dot{\hat{c}}_0=-i\bar{c}_-(g_a\hat{a}+g_b\hat{b})-i\bar{c}_+(g_a\hat{a}^\dag+g_b\hat{b}^\dag)-\frac{\kappa}{2}\hat{c}_0.\\
    \end{split}
\end{equation}
This set of equations is exactly equivalent to the dynamics governed by an effective Hamiltonian:

\begin{equation}
    \label{eq:H3mode}
    \mathcal{H}=\hbar\Omega(\hat{a}^\dag\hat{a}-\hat{b}^\dag\hat{b})+\hbar k_M(\hat{a}^\dag\hat{b}+\hat{b}^\dag\hat{a})+\hbar g_a[(\bar{c}_-\hat{a}+\bar{c}_+\hat{a}^\dag)\hat{c}_0^\dag+\mathrm{H.c.}]+g_b[(\bar{c}_-\hat{b}+\bar{c}_+\hat{b}^\dag)\hat{c}_0^\dag+\mathrm{H.c.}].
\end{equation}

Under the adiabatic approximation, we can assume that the cavity responds rapidly to the mechanical motion, so $\hat{c}_0$ can be expressed in terms of $\hat a$ and $\hat b$ using the third equation in Eq.~\eqref{eq:HL4}:
\begin{equation}
\label{eq:c0}
\hat{c}_0\approx -i\frac{2}{\kappa}[\bar{c}_-(g_a\hat{a}+g_b\hat{b})+\bar{c}_+(g_a\hat{a}^\dag+g_b\hat{b}^\dag)].
\end{equation}

Substituting Eq.~\eqref{eq:c0} into Eq.~\eqref{eq:H3mode}, the single-photon optomechanical coupling terms in the Hamiltonian are canceled, yielding the effective two-mode Hamiltonian:

\begin{equation}    \mathcal{H}_{\mathrm{eff}}=\hbar\Omega(\hat{a}^\dag\hat{a}-\hat{b}^\dag\hat{b})+\hbar k_M(\hat{a}^\dag\hat{b}+\hat{b}^\dag\hat{a}).
\end{equation}

Based on this effective Hamiltonian, we can write down the effective two-mode master equation:

\begin{equation}
    \begin{split}
        \dot{\hat{\rho}} =& -\frac{i}{\hbar}[\hbar \Omega (\hat{a}^\dagger \hat{a} - \hat{b}^\dagger \hat{b}) + \hbar k_M (\hat{a}^\dagger \hat{b} + \hat{a}\hat{b}^\dagger), \hat{\rho}]+ \gamma_a (\overline{n}_a + 1) \mathcal{D}[\hat{a}]\hat{\rho} + \gamma_a \overline{n}_a \mathcal{D}[\hat{a}^\dagger]\hat{\rho}+ \gamma_b (\overline{n}_b + 1) \mathcal{D}[\hat{b}]\hat{\rho} + \gamma_b \overline{n}_b \mathcal{D}[\hat{b}^\dagger]\hat{\rho}\\
        &+ \kappa \mathcal{D}[\hat{c}_0]\hat{\rho}+ 2(\kappa_{G,a}\mathcal{D}[\hat{a}]\hat{\rho}+ \kappa_{G,a}\mathcal{D}[\hat{a}^\dagger]\hat{\rho} + \kappa_{G,b}\mathcal{D}[\hat{b}]\hat{\rho}+ \kappa_{G,b}\mathcal{D}[\hat{b}^\dagger]\hat{\rho}).
        \label{eq:me2}
    \end{split}
\end{equation}
Here, the dissipative terms can be carried over directly from Eq.~\eqref{eq:me} . Note that the dissipation for $\hat{c}$ is replaced by that for $\hat{c}_0$, since we have already neglected the high-frequency components.

In the main text, we introduced the operators $\hat{\beta}_1$ and $\hat{\beta}_2$. From their and other parameters' definitions, we have:

\begin{equation}
    \begin{split}
        &\hat{\beta}_1=\hat{a}\cosh{r}+\hat{b}^\dag\sinh{r}=\hat{a}\frac{1}{\sqrt{1-(G_+/G_-)^2}}+\hat{b}^\dag\frac{G_+/G_-}{\sqrt{1-(G_+/G_-)^2}}=\hat{a}\frac{G_-}{\mathcal{G}}+\hat{b}^\dag\frac{G_+}{\mathcal{G}}=\frac{g_a+g_b}{2\mathcal{G}}(\bar{c}_-\hat{a}+\bar{c}_+\hat{b}^\dag),\\
        &\hat{\beta}_2=\hat{b}\cosh{r}+\hat{a}^\dag\sinh{r}=\hat{b}\frac{1}{\sqrt{1-(G_+/G_-)^2}}+\hat{a}^\dag\frac{G_+/G_-}{\sqrt{1-(G_+/G_-)^2}}=\hat{b}\frac{G_-}{\mathcal{G}}+\hat{a}^\dag\frac{G_+}{\mathcal{G}}=\frac{g_a+g_b}{2\mathcal{G}}(\bar{c}_-\hat{b}+\bar{c}_+\hat{a}^\dag).
        \label{eq:beta}
    \end{split}
\end{equation}
When the single-photon coupling strengths $g_a$ and $g_b$ are equal, we can find through Eq.~\eqref{eq:c0} and Eq.~\eqref{eq:beta} that:

\begin{equation}
    \hat{c}_0\approx \frac{-2i\mathcal{G}(\hat{\beta}_1+\hat{\beta}_2)}{\kappa}.
\end{equation}
Substituting into Eq.~\eqref{eq:me2} , we obtain the final effective two-mode master equation as in the main text: 

\begin{equation}
    \begin{split}
            \dot{\hat{\rho}} =& -\frac{i}{\hbar}[\hbar \Omega (\hat{a}^\dagger \hat{a} - \hat{b}^\dagger \hat{b}) + \hbar k_M (\hat{a}^\dagger \hat{b} + \hat{a}\hat{b}^\dagger), \hat{\rho}]+ \gamma_a (\overline{n}_a + 1) \mathcal{D}[\hat{a}]\hat{\rho} + \gamma_a \overline{n}_a \mathcal{D}[\hat{a}^\dagger]\hat{\rho}+ \gamma_b (\overline{n}_b + 1) \mathcal{D}[\hat{b}]\hat{\rho} + \gamma_b \overline{n}_b \mathcal{D}[\hat{b}^\dagger]\hat{\rho}\\
        &+ \Gamma \mathcal{D}[\hat{\beta}_1+\hat{\beta}_2]\hat{\rho}+ 2(\kappa_{G,a}\mathcal{D}[\hat{a}]\hat{\rho}+ \kappa_{G,a}\mathcal{D}[\hat{a}^\dagger]\hat{\rho} + \kappa_{G,b}\mathcal{D}[\hat{b}]\hat{\rho}+ \kappa_{G,b}\mathcal{D}[\hat{b}^\dagger]\hat{\rho})
        \label{eq:me3}
    \end{split} 
\end{equation}
where $\Gamma=4\mathcal{G}^2/\kappa$. This equation serves as the basis for further estimating the entanglement between the two mechanical oscillators.

\section{Derivation of the Entanglement}

In this section, we derive one of the core results of the main text—the expression for the entanglement between the two mechanical oscillators. We have set $\gamma_a=\gamma_b=\gamma_m$, $\bar{n}_a=\bar{n}_b=\bar{n}$ and $\kappa_{G,a}=\kappa_{G,a}=\kappa_{G}$. These symmetries are imposed solely to simplify the expressions. Results for the more general case can be obtained by following the same derivation.

We can use $\frac{d\langle \hat{O}\rangle}{dt}=\mathrm{Tr}\!\left[\hat{O}\,\dot{\hat{\rho}}\right]$ and, by substituting Eq.~\eqref{eq:me3}, obtain the dynamical equation of any second-order moment $\langle \hat{O}\rangle$. Further assuming the system reaches a steady state, we obtain the following set of equations (although some second-order moments are complex conjugates of each other, we treat them as independent, thereby formulating a 10-variable linear equation ):

\begin{subequations}\renewcommand{\theequation}{\theparentequation\Alph{equation}}
\begin{equation}
(\gamma_m +\Gamma +2 i \Omega ) \langle \hat{a}^2 \rangle+\left(\Gamma +2 i k_M\right) \langle \hat{a}\hat{b} \rangle +\frac{1}{2} \Gamma  \sinh (2 r) =0,
\end{equation}
\begin{equation}
(\gamma_m +\Gamma -2 i \Omega ) \langle \hat{b}^2 \rangle +\left(\Gamma +2 i k_M\right) \langle \hat{a}\hat{b} \rangle +\frac{1}{2} \Gamma  \sinh (2 r) =0, 
\end{equation}
\begin{equation}
(\gamma_m +\Gamma -2 i \Omega ) \langle \hat{a}^{\dagger2} \rangle +\left(\Gamma -2 i k_M\right) \langle \hat{a}^\dagger\hat{b}^\dagger \rangle +\frac{1}{2} \Gamma  \sinh (2 r) =0, 
\end{equation}
\begin{equation}
(\gamma_m +\Gamma +2 i \Omega ) \langle \hat{b}^{\dagger2}  \rangle +\left(\Gamma -2 i k_M\right) \langle \hat{a}^\dagger\hat{b}^\dagger \rangle +\frac{1}{2} \Gamma  \sinh (2 r) =0, 
\end{equation}
\begin{equation}
(\gamma_m +\Gamma +2 i \Omega ) \langle \hat{a}\hat{b}^\dagger \rangle +\left(\frac{\Gamma }{2}-i k_M\right) \langle \hat{a}^\dagger\hat{a} \rangle +\left(\frac{\Gamma }{2}+i k_M\right) \langle \hat{b}^\dagger\hat{b} \rangle +\Gamma  \sinh ^2(r) =0, 
\end{equation}
\begin{equation}
(\gamma_m +\Gamma -2 i \Omega ) \langle \hat{a}^\dagger\hat{b} \rangle +\left(\frac{\Gamma }{2}+i k_M\right) \langle \hat{a}^\dagger\hat{a} \rangle +\left(\frac{\Gamma }{2}-i k_M\right)
   \langle \hat{b}^\dagger\hat{b} \rangle -\Gamma  \sinh ^2(r) =0, 
   \end{equation}
\begin{equation}
(\gamma_m +\Gamma ) \langle \hat{a}^\dagger\hat{a} \rangle +\left(\frac{\Gamma }{2}+i k_M\right) \langle \hat{a}^\dagger\hat{b} \rangle +\left(\frac{\Gamma }{2}-i k_M\right) \langle \hat{a}\hat{b}^\dagger \rangle -\left(\Gamma 
   \sinh ^2(r)+\bar{n} \gamma_m +2 \kappa _G\right) =0, 
   \end{equation}
\begin{equation}
(\gamma_m +\Gamma ) \langle \hat{a}\hat{b} \rangle +\left(\frac{\Gamma }{2}+i k_M\right) \left(\langle \hat{a}^2 \rangle + \langle \hat{b}^2 \rangle \right)+\frac{1}{2} \Gamma  \sinh (2 r) =0, 
\end{equation}
\begin{equation}
(\gamma_m +\Gamma ) \langle \hat{b}^\dagger\hat{b} \rangle +\left(\frac{\Gamma }{2}+i k_M\right) \langle \hat{a}\hat{b}^\dagger \rangle +\left(\frac{\Gamma }{2}-i k_M\right) \langle \hat{a}^\dagger\hat{b} \rangle -\left(\Gamma 
   \sinh ^2(r)+\bar{n} \gamma_m +2 \kappa _G\right) =0, 
   \end{equation}
\begin{equation}
(\gamma_m +\Gamma ) \langle \hat{a}^\dagger\hat{b}^\dagger \rangle +\left(\frac{\Gamma }{2}-i k_M\right) \left(\langle \hat{a}^{\dagger2} \rangle +\langle \hat{b}^{\dagger2}\rangle\right)+\frac{1}{2} \Gamma  \sinh (2 r) =0. 
\end{equation}
\end{subequations}

Solving this linear equation for the second-order moments, we obtain:

\begin{subequations}\renewcommand{\theequation}{\theparentequation\Alph{equation}}
\label{eq:second_moment_ab}
\begin{equation}
  \langle \hat{a}^2 \rangle  = \frac{(\gamma_m +\Gamma -2 i \Omega ) \left(\gamma_m -2 i k_M\right) \Gamma  \sinh (2 r)}{2 (\gamma_m +\Gamma ) \left(\Gamma ^2+4 i k_M \Gamma -A\right)}, 
  \end{equation}
\begin{equation}
  \langle \hat{b}^2 \rangle  = \frac{(\gamma_m +\Gamma +2 i \Omega ) \left(\gamma_m -2 i k_M\right) \Gamma  \sinh (2 r)}{2 (\gamma_m +\Gamma ) \left(\Gamma ^2+4 i k_M \Gamma -A\right)}, 
  \end{equation}
\begin{equation}
  \langle \hat{a}^{\dagger2} \rangle  = \frac{(\gamma_m +\Gamma +2 i \Omega ) \left(\gamma_m +2 i k_M\right) \Gamma  \sinh (2 r)}{2 (\gamma_m +\Gamma ) \left(\Gamma ^2-4 i k_M \Gamma -A\right)}, 
  \end{equation}
\begin{equation}
  \langle \hat{b}^{\dagger2} \rangle  = \frac{(\gamma_m +\Gamma -2 i \Omega ) \left(\gamma_m +2 i k_M\right) \Gamma  \sinh (2 r)}{2 (\gamma_m +\Gamma ) \left(\Gamma ^2-4 i k_M \Gamma -A\right)}, 
  \end{equation}
\begin{equation}
  \langle \hat{a} \hat{b}^\dagger \rangle  = -\frac{\Gamma  \left((\gamma_m +\Gamma ) (\gamma_m +\Gamma -2 i \Omega )+4 k_M^2\right) \left(\left(\bar{n}-\sinh ^2(r)\right) \gamma_m +2 \kappa _G\right)}{\gamma_m  (\gamma_m +2 \Gamma ) A+4 \Gamma ^2 \Omega ^2}, 
  \end{equation}
\begin{equation}
  \langle \hat{a}^\dagger \hat{b} \rangle  = -\frac{\Gamma  \left((\gamma_m +\Gamma ) (\gamma_m +\Gamma +2 i \Omega )+4 k_M^2\right) \left(\left(\bar{n}-\sinh ^2(r)\right) \gamma_m +2 \kappa _G\right)}{\gamma_m  (\gamma_m +2 \Gamma ) A+4 \Gamma ^2 \Omega ^2}, 
  \end{equation}
\begin{equation}
  \langle \hat{a}^\dagger \hat{a} \rangle  = \frac{\left(\bar{n} \gamma_m +2 \kappa _G\right) \left((\gamma_m +\Gamma ) A-4 \Gamma  \Omega  k_M\right)+\left(\gamma_m  A+4 \Omega  \left(\Gamma  \Omega +k_M \gamma_m \right)\right) \Gamma \sinh ^2(r)}{\gamma_m  (\gamma_m +2 \Gamma ) A+4 \Gamma ^2 \Omega ^2}, 
  \end{equation}
\begin{equation}
  \langle \hat{a} \hat{b} \rangle  = \frac{\left((\gamma_m +\Gamma ) \left(\gamma_m -2 i k_M\right)+4 \Omega ^2\right) \Gamma  \sinh (2 r)}{2 (\gamma_m +\Gamma ) \left(\Gamma ^2+4 i k_M \Gamma -A\right)}, 
  \end{equation}
\begin{equation}
  \langle \hat{b}^\dagger \hat{b} \rangle  = \frac{\left(\bar{n} \gamma_m +2 \kappa _G\right) \left((\gamma_m +\Gamma ) A+4 \Gamma  \Omega  k_M\right)+\left(\gamma_m  A+4 \Omega  \left(\Gamma  \Omega -k_M \gamma_m \right)\right) \Gamma \sinh ^2(r)}{\gamma_m  (\gamma_m +2 \Gamma ) A+4 \Gamma ^2 \Omega ^2}, 
  \end{equation}
\begin{equation}
  \langle \hat{a}^\dagger \hat{b}^\dagger \rangle  = \frac{\left((\gamma_m +\Gamma ) \left(\gamma_m +2 i k_M\right)+4 \Omega ^2\right) \Gamma  \sinh (2 r)}{2 (\gamma_m +\Gamma ) \left(\Gamma ^2-4 i k_M \Gamma -A\right)}
\end{equation}
\end{subequations}
where $A = (\gamma_m +\Gamma )^2+4 \left(\Omega ^2+k_M^2\right)$.

The covariance matrix between the two oscillators $a$ and $b$ is defined as \cite{bowen2015quantum,aspelmeyer2014cavity}:

\begin{equation}
\label{eq:def_cov}
V=
\begin{pmatrix}
V_a & V_{ab} \\
V_{ab}^T & V_b
\end{pmatrix}
=
\begin{pmatrix}
V_{\hat{X}_a\hat{X}_a} & V_{\hat{X}_a\hat{P}_a} & V_{\hat{X}_a\hat{X}_b} & V_{\hat{X}_a\hat{P}_b} \\
V_{\hat{P}_a\hat{X}_a} & V_{\hat{P}_a\hat{P}_a} & V_{\hat{P}_a\hat{X}_b} & V_{\hat{P}_a\hat{P}_b} \\
V_{\hat{X}_b\hat{X}_a} & V_{\hat{X}_b\hat{P}_a} & V_{\hat{X}_b\hat{X}_b} & V_{\hat{X}_b\hat{P}_b} \\
V_{\hat{P}_b\hat{X}_a} & V_{\hat{P}_a\hat{P}_a} & V_{\hat{P}_b\hat{X}_b} & V_{\hat{P}_a\hat{P}_b}
\end{pmatrix}
\end{equation}
where $V_{AB}=\frac{\langle AB \rangle+\langle BA \rangle}{2}-\langle A \rangle \langle B \rangle$. $V_a$, $V_b$ and $V_{ab}$ are $2\times 2$ sub-matrices. The covariance matrix is symmetric and all matrix elements can be calculated from the expressions in Eq.~\eqref{eq:second_moment_ab} using the relations $\hat{X}_a=(\hat{a}+\hat{a}^\dag)/\sqrt{2}$, $\hat{X}_b=(\hat{b}+\hat{b}^\dag)/\sqrt{2}$, $\hat{P}_a=-i(\hat{a}-\hat{a}^\dag)/\sqrt{2}$ and $\hat{P}_b=-i(\hat{b}-\hat{b}^\dag)/\sqrt{2}$. The results are as follows, where $\alpha=k_M/\Omega$:

\begin{subequations}\renewcommand{\theequation}{\theparentequation\Alph{equation}}

\begin{equation}
\begin{split}
V_{\hat{X}_a\hat{X}_a}=\langle \hat{X}_a^2 \rangle = & \frac{1}{4} \bigg( \Big[ 4 \big(4 \Omega ^2 (2 \gamma_m  \bar{n} ((\alpha ^2+1) \gamma_m +(\alpha -1) \alpha  \Gamma +\Gamma )-\Gamma  ((\alpha ^2+\alpha +1) \gamma_m +\Gamma )) \\
    & \qquad + \gamma_m  (\gamma_m +\Gamma )^2 (2 \bar{n} (\gamma_m +\Gamma )-\Gamma )+4 \kappa _G (4 \Omega ^2 ((\alpha ^2+1) \gamma_m +(\alpha -1) \alpha  \Gamma +\Gamma )+(\gamma_m +\Gamma )^3) \\
    & \qquad + \Gamma  \cosh (2 r) (4 \Omega ^2 ((\alpha ^2+\alpha +1) \gamma_m +\Gamma )+\gamma_m  (\gamma_m +\Gamma )^2) \big) \Big] \\
    & \qquad \times \Big[ 8 \Omega ^2 ((\alpha ^2+1) \gamma_m ^2+2 (\alpha ^2+1) \gamma_m  \Gamma +\Gamma ^2)+2 \gamma_m  (\gamma_m +2 \Gamma ) (\gamma_m +\Gamma )^2 \Big]^{-1} \\
    & \qquad -\frac{\Gamma  \sinh (2 r) (\gamma_m -2 i \alpha  \Omega ) (\gamma_m +\Gamma -2 i \Omega )}{(\gamma_m +\Gamma ) (4 \Omega  (\alpha ^2 \Omega -i \alpha  \Gamma +\Omega )+\gamma_m ^2+2 \gamma_m  \Gamma )} \\
    & \qquad -\frac{\Gamma  \sinh (2 r) (\gamma_m +2 i \alpha  \Omega ) (\gamma_m +\Gamma +2 i \Omega )}{(\gamma_m +\Gamma ) (4 \Omega  (\alpha ^2 \Omega +i \alpha  \Gamma +\Omega )+\gamma_m ^2+2 \gamma_m  \Gamma )}+2 \bigg),
\end{split}
\end{equation}

\begin{equation}
\begin{split}
	V_{\hat{X}_a\hat{P}_a}=\frac12 \langle \hat{X}_a \hat{P}_a +  \hat{P}_a \hat{X}_a\rangle = & -\Gamma \Omega \sinh (2 r) \Big(4 (\alpha ^3+\alpha ^2+\alpha +1) \gamma_m \Omega ^2+(\alpha +1) \gamma_m ^3 +(\alpha +2) \gamma_m ^2 \Gamma \\
	& + 4 \alpha (\alpha +1)^2 \Gamma \Omega ^2\Big) \times \Big[(\gamma_m +\Gamma ) \left(4 \Omega (\alpha ^2 \Omega -i \alpha \Gamma +\Omega )+\gamma_m ^2+2 \gamma_m \Gamma \right) \\
	& \times \left(4 \Omega (\alpha ^2 \Omega +i \alpha \Gamma +\Omega )+\gamma_m ^2+2 \gamma_m \Gamma \right) \Big]^{-1},
\end{split}
\end{equation}

\begin{equation}
\begin{split}
	V_{\hat{X}_a\hat{X}_b}=\frac12 \langle \hat{X}_a \hat{X}_b + \hat{X}_b \hat{X}_a\rangle = & \frac{1}{4} \Gamma \bigg( -2 \Big[ (4 \alpha ^2 \Omega ^2+2 i \Omega (\gamma_m +\Gamma )+(\gamma_m +\Gamma )^2) (2 \gamma_m \bar{n}+\gamma_m +4 \kappa _G-\gamma_m \cosh (2 r)) \\
	& \qquad + (4 \alpha ^2 \Omega ^2-2 i \Omega (\gamma_m +\Gamma )+(\gamma_m +\Gamma )^2) (2 \gamma_m \bar{n}+\gamma_m +4 \kappa _G-\gamma_m \cosh (2 r)) \Big] \\
	& \qquad \times \Big[ 8 \Omega ^2 ((\alpha ^2+1) \gamma_m ^2+2 (\alpha ^2+1) \gamma_m \Gamma +\Gamma ^2)+2 \gamma_m (\gamma_m +2 \Gamma ) (\gamma_m +\Gamma )^2 \Big]^{-1} \\
	& \qquad -\frac{\sinh (2 r) (-2 i \alpha \Omega (\gamma_m +\Gamma )+\gamma_m (\gamma_m +\Gamma )+4 \Omega ^2)}{(\gamma_m +\Gamma ) (4 \Omega (\alpha ^2 \Omega -i \alpha \Gamma +\Omega )+\gamma_m ^2+2 \gamma_m \Gamma )} \\
	& \qquad -\frac{\sinh (2 r) (2 i \alpha \Omega (\gamma_m +\Gamma )+\gamma_m (\gamma_m +\Gamma )+4 \Omega ^2)}{(\gamma_m +\Gamma ) (4 \Omega (\alpha ^2 \Omega +i \alpha \Gamma +\Omega )+\gamma_m ^2+2 \gamma_m \Gamma )} \bigg),
\end{split}
\end{equation}

\begin{equation}
\begin{split}
	V_{\hat{X}_a\hat{P}_b}=\frac12 \langle \hat{X}_a \hat{P}_b + \hat{P}_b \hat{X}_a\rangle= & -\frac{1}{4} i \Gamma \bigg( 2 \Big[ (4 \alpha ^2 \Omega ^2+2 i \Omega (\gamma_m +\Gamma )+(\gamma_m +\Gamma )^2) (2 \gamma_m \bar{n}+\gamma_m +4 \kappa _G-\gamma_m \cosh (2 r)) \\
	& \qquad - (4 \alpha ^2 \Omega ^2-2 i \Omega (\gamma_m +\Gamma )+(\gamma_m +\Gamma )^2) (2 \gamma_m \bar{n}+\gamma_m +4 \kappa _G-\gamma_m \cosh (2 r)) \Big] \\
	& \qquad \times \Big[ 8 \Omega ^2 ((\alpha ^2+1) \gamma_m ^2+2 (\alpha ^2+1) \gamma_m \Gamma +\Gamma ^2)+2 \gamma_m (\gamma_m +2 \Gamma ) (\gamma_m +\Gamma )^2 \Big]^{-1} \\
	& \qquad +\frac{\sinh (2 r) (-2 i \alpha \Omega (\gamma_m +\Gamma )+\gamma_m (\gamma_m +\Gamma )+4 \Omega ^2)}{(\gamma_m +\Gamma ) (4 \Omega (\alpha ^2 \Omega -i \alpha \Gamma +\Omega )+\gamma_m ^2+2 \gamma_m \Gamma )} \\
	& \qquad -\frac{\sinh (2 r) (2 i \alpha \Omega (\gamma_m +\Gamma )+\gamma_m (\gamma_m +\Gamma )+4 \Omega ^2)}{(\gamma_m +\Gamma ) (4 \Omega (\alpha ^2 \Omega +i \alpha \Gamma +\Omega )+\gamma_m ^2+2 \gamma_m \Gamma )} \bigg),
\end{split}
\end{equation}

\begin{equation}
\begin{split}
	V_{\hat{P}_a\hat{P}_a}=\langle \hat{P}_a^2 \rangle = & \frac{1}{4} \bigg( \Big[ 4 \big(4 \Omega ^2 (2 \gamma_m  \bar{n} ((\alpha ^2+1) \gamma_m +(\alpha -1) \alpha  \Gamma +\Gamma )-\Gamma  ((\alpha ^2+\alpha +1) \gamma_m +\Gamma )) \\
	& \qquad + \gamma_m  (\gamma_m +\Gamma )^2 (2 \bar{n} (\gamma_m +\Gamma )-\Gamma )+4 \kappa _G (4 \Omega ^2 ((\alpha ^2+1) \gamma_m +(\alpha -1) \alpha  \Gamma +\Gamma )+(\gamma_m +\Gamma )^3) \\
	& \qquad + \Gamma  \cosh (2 r) (4 \Omega ^2 ((\alpha ^2+\alpha +1) \gamma_m +\Gamma )+\gamma_m  (\gamma_m +\Gamma )^2) \big) \Big] \\
	& \qquad \times \Big[ 8 \Omega ^2 ((\alpha ^2+1) \gamma_m ^2+2 (\alpha ^2+1) \gamma_m  \Gamma +\Gamma ^2)+2 \gamma_m  (\gamma_m +2 \Gamma ) (\gamma_m +\Gamma )^2 \Big]^{-1} \\
	& \qquad +\frac{\Gamma  \sinh (2 r) (\gamma_m -2 i \alpha  \Omega ) (\gamma_m +\Gamma -2 i \Omega )}{(\gamma_m +\Gamma ) (4 \Omega  (\alpha ^2 \Omega -i \alpha  \Gamma +\Omega )+\gamma_m ^2+2 \gamma_m  \Gamma )} \\
	& \qquad +\frac{\Gamma  \sinh (2 r) (\gamma_m +2 i \alpha  \Omega ) (\gamma_m +\Gamma +2 i \Omega )}{(\gamma_m +\Gamma ) (4 \Omega  (\alpha ^2 \Omega +i \alpha  \Gamma +\Omega )+\gamma_m ^2+2 \gamma_m  \Gamma )}+2 \bigg),
\end{split}
\end{equation}

\begin{equation}
\begin{split}
	V_{\hat{X}_b\hat{P}_a}=\frac12  \langle \hat{X}_b \hat{P}_a +  \hat{P}_a \hat{X}_b\rangle = & -\frac{1}{4} i \Gamma \bigg( 2 \Big[ (4 \alpha ^2 \Omega ^2-2 i \Omega (\gamma_m +\Gamma )+(\gamma_m +\Gamma )^2) (2 \gamma_m \bar{n}+\gamma_m +4 \kappa _G-\gamma_m \cosh (2 r)) \\
	& \qquad - (4 \alpha ^2 \Omega ^2+2 i \Omega (\gamma_m +\Gamma )+(\gamma_m +\Gamma )^2) (2 \gamma_m \bar{n}+\gamma_m +4 \kappa _G-\gamma_m \cosh (2 r)) \Big] \\
	& \qquad \times \Big[ 8 \Omega ^2 ((\alpha ^2+1) \gamma_m ^2+2 (\alpha ^2+1) \gamma_m \Gamma +\Gamma ^2)+2 \gamma_m (\gamma_m +2 \Gamma ) (\gamma_m +\Gamma )^2 \Big]^{-1} \\
	& \qquad +\frac{\sinh (2 r) (-2 i \alpha \Omega (\gamma_m +\Gamma )+\gamma_m (\gamma_m +\Gamma )+4 \Omega ^2)}{(\gamma_m +\Gamma ) (4 \Omega (\alpha ^2 \Omega -i \alpha \Gamma +\Omega )+\gamma_m ^2+2 \gamma_m \Gamma )} \\
	& \qquad -\frac{\sinh (2 r) (2 i \alpha \Omega (\gamma_m +\Gamma )+\gamma_m (\gamma_m +\Gamma )+4 \Omega ^2)}{(\gamma_m +\Gamma ) (4 \Omega (\alpha ^2 \Omega +i \alpha \Gamma +\Omega )+\gamma_m ^2+2 \gamma_m \Gamma )} \bigg),
\end{split}
\end{equation}

\begin{equation}
\begin{split}
	V_{\hat{P}_a\hat{P}_b}=\frac12 \langle \hat{P}_a \hat{P}_b + \hat{P}_b \hat{P}_a\rangle = & \frac{1}{4} \Gamma \bigg( -2 \Big[ (4 \alpha ^2 \Omega ^2+2 i \Omega (\gamma_m +\Gamma )+(\gamma_m +\Gamma )^2) (2 \gamma_m \bar{n}+\gamma_m +4 \kappa _G-\gamma_m \cosh (2 r)) \\
	& \qquad + (4 \alpha ^2 \Omega ^2-2 i \Omega (\gamma_m +\Gamma )+(\gamma_m +\Gamma )^2) (2 \gamma_m \bar{n}+\gamma_m +4 \kappa _G-\gamma_m \cosh (2 r)) \Big] \\
	& \qquad \times \Big[ 8 \Omega ^2 ((\alpha ^2+1) \gamma_m ^2+2 (\alpha ^2+1) \gamma_m \Gamma +\Gamma ^2)+2 \gamma_m (\gamma_m +2 \Gamma ) (\gamma_m +\Gamma )^2 \Big]^{-1} \\
	& \qquad +\frac{\sinh (2 r) (-2 i \alpha \Omega (\gamma_m +\Gamma )+\gamma_m (\gamma_m +\Gamma )+4 \Omega ^2)}{(\gamma_m +\Gamma ) (4 \Omega (\alpha ^2 \Omega -i \alpha \Gamma +\Omega )+\gamma_m ^2+2 \gamma_m \Gamma )} \\
	& \qquad +\frac{\sinh (2 r) (2 i \alpha \Omega (\gamma_m +\Gamma )+\gamma_m (\gamma_m +\Gamma )+4 \Omega ^2)}{(\gamma_m +\Gamma ) (4 \Omega (\alpha ^2 \Omega +i \alpha \Gamma +\Omega )+\gamma_m ^2+2 \gamma_m \Gamma )} \bigg),
\end{split}
\end{equation}

\begin{equation}
\begin{split}
	V_{\hat{X}_b\hat{X}_b}=\langle \hat{X}_b^2 \rangle = & \frac{1}{4} \bigg( \Big[ 4 \big(4 \Omega ^2 (2 \gamma_m  \bar{n} ((\alpha ^2+1) \gamma_m +\alpha (\alpha +1) \Gamma +\Gamma )-\Gamma ((\alpha -1) \alpha \gamma_m +\gamma_m +\Gamma )) \\
	& \qquad + \gamma_m (\gamma_m +\Gamma )^2 (2 \bar{n} (\gamma_m +\Gamma )-\Gamma )+4 \kappa _G (4 \Omega ^2 ((\alpha ^2+1) \gamma_m +\alpha (\alpha +1) \Gamma +\Gamma )+(\gamma_m +\Gamma )^3) \\
	& \qquad + \Gamma \cosh (2 r) (4 \Omega ^2 ((\alpha -1) \alpha \gamma_m +\gamma_m +\Gamma )+\gamma_m (\gamma_m +\Gamma )^2) \big) \Big] \\
	& \qquad \times \Big[ 8 \Omega ^2 ((\alpha ^2+1) \gamma_m ^2+2 (\alpha ^2+1) \gamma_m \Gamma +\Gamma ^2)+2 \gamma_m (\gamma_m +2 \Gamma ) (\gamma_m +\Gamma )^2 \Big]^{-1} \\
	& \qquad -\frac{\Gamma \sinh (2 r) (\gamma_m -2 i \alpha \Omega ) (\gamma_m +\Gamma +2 i \Omega )}{(\gamma_m +\Gamma ) (4 \Omega (\alpha ^2 \Omega -i \alpha \Gamma +\Omega )+\gamma_m ^2+2 \gamma_m \Gamma )} \\
	& \qquad -\frac{\Gamma \sinh (2 r) (\gamma_m +2 i \alpha \Omega ) (\gamma_m +\Gamma -2 i \Omega )}{(\gamma_m +\Gamma ) (4 \Omega (\alpha ^2 \Omega +i \alpha \Gamma +\Omega )+\gamma_m ^2+2 \gamma_m \Gamma )} + 2 \bigg),
\end{split}
\end{equation}

\begin{equation}
\begin{split}
	V_{\hat{X}_b\hat{P}_b}=\frac12 \langle \hat{X}_b \hat{P}_b + \hat{P}_b \hat{X}_b\rangle = & -\Gamma \Omega \sinh (2 r) \Big(4 (\alpha ^3-\alpha ^2+\alpha -1) \gamma_m \Omega ^2+(\alpha -1) \gamma_m ^3+(\alpha -2) \gamma_m ^2 \Gamma \\
	& + 4 (\alpha -1)^2 \alpha \Gamma \Omega ^2\Big) \times \Big[(\gamma_m +\Gamma ) \left(4 \Omega (\alpha ^2 \Omega -i \alpha \Gamma +\Omega )+\gamma_m ^2+2 \gamma_m \Gamma \right) \\
	& \times \left(4 \Omega (\alpha ^2 \Omega +i \alpha \Gamma +\Omega )+\gamma_m ^2+2 \gamma_m \Gamma \right) \Big]^{-1},
\end{split}
\end{equation}

\begin{equation}
\begin{split}
	V_{\hat{P}_b\hat{P}_b}=\langle \hat{P}_b^2 \rangle = & \frac{1}{4} \bigg( \Big[ 4 \big(4 \Omega ^2 (2 \gamma_m  \bar{n} ((\alpha ^2+1) \gamma_m +\alpha (\alpha +1) \Gamma +\Gamma )-\Gamma ((\alpha -1) \alpha \gamma_m +\gamma_m +\Gamma )) \\
	& \qquad + \gamma_m (\gamma_m +\Gamma )^2 (2 \bar{n} (\gamma_m +\Gamma )-\Gamma )+4 \kappa _G (4 \Omega ^2 ((\alpha ^2+1) \gamma_m +\alpha (\alpha +1) \Gamma +\Gamma )+(\gamma_m +\Gamma )^3) \\
	& \qquad + \Gamma \cosh (2 r) (4 \Omega ^2 ((\alpha -1) \alpha \gamma_m +\gamma_m +\Gamma )+\gamma_m (\gamma_m +\Gamma )^2) \big) \Big] \\
	& \qquad \times \Big[ 8 \Omega ^2 ((\alpha ^2+1) \gamma_m ^2+2 (\alpha ^2+1) \gamma_m \Gamma +\Gamma ^2)+2 \gamma_m (\gamma_m +2 \Gamma ) (\gamma_m +\Gamma )^2 \Big]^{-1} \\
	& \qquad +\frac{\Gamma \sinh (2 r) (\gamma_m -2 i \alpha \Omega ) (\gamma_m +\Gamma +2 i \Omega )}{(\gamma_m +\Gamma ) (4 \Omega (\alpha ^2 \Omega -i \alpha \Gamma +\Omega )+\gamma_m ^2+2 \gamma_m \Gamma )} \\
	& \qquad +\frac{\Gamma \sinh (2 r) (\gamma_m +2 i \alpha \Omega ) (\gamma_m +\Gamma -2 i \Omega )}{(\gamma_m +\Gamma ) (4 \Omega (\alpha ^2 \Omega +i \alpha \Gamma +\Omega )+\gamma_m ^2+2 \gamma_m \Gamma )} + 2 \bigg).
\end{split}
\end{equation}
\end{subequations}
Take the limit that the mechanical dissipation is relatively small, $\Omega/\gamma_m\rightarrow \infty$, we have:
\begin{subequations}\renewcommand{\theequation}{\theparentequation\Alph{equation}}
\begin{equation}\begin{split}
	\langle \hat{X}_a^2 \rangle \approx & \Big( 2 \bar{n} (\alpha^4 \gamma_m^3 + 2\alpha^4 \gamma_m^2 \Gamma + \alpha^4 \gamma_m \Gamma^2 - \alpha^3 \gamma_m^2 \Gamma - \alpha^3 \gamma_m \Gamma^2 + 2\alpha^2 \gamma_m^3 + 4\alpha^2 \gamma_m^2 \Gamma + 2\alpha^2 \gamma_m \Gamma^2 \\
	& \qquad - \alpha \gamma_m^2 \Gamma - \alpha \gamma_m \Gamma^2 + \gamma_m^3 + 2\gamma_m^2 \Gamma + \gamma_m \Gamma^2) + (\alpha^4 \gamma_m^3 + 2\alpha^4 \gamma_m^2 \Gamma + \alpha^4 \gamma_m \Gamma^2 \\
	& \qquad - \alpha^3 \gamma_m^2 \Gamma - \alpha^3 \gamma_m \Gamma^2 + 2\alpha^2 \gamma_m^3 + 4\alpha^2 \gamma_m^2 \Gamma + 2\alpha^2 \gamma_m \Gamma^2 - \alpha \gamma_m^2 \Gamma - \alpha \gamma_m \Gamma^2 + \gamma_m^3 + 2\gamma_m^2 \Gamma + \gamma_m \Gamma^2) \\
	& \qquad + 4 (\alpha^2+1) (\gamma_m +\Gamma ) \kappa _G (\alpha^2 \gamma_m +\alpha^2 \Gamma -\alpha \Gamma +\gamma_m +\Gamma ) \\
	& \qquad + \sinh(2r) (\alpha^3 \gamma_m^2 \Gamma + 2\alpha^3 \gamma_m \Gamma^2 + \alpha \gamma_m^2 \Gamma + 2\alpha \gamma_m \Gamma^2 + \alpha \Gamma^3) \\
	& \qquad + (\alpha^2+1) \Gamma (\gamma_m +\Gamma ) \cosh (2 r) ((\alpha^2+\alpha +1) \gamma_m +\Gamma ) \Big) \times D,
\end{split}\end{equation}
\begin{equation}
	\frac{1}{2}\langle \hat{X}_a \hat{P}_a +   \hat{P}_a \hat{X}_a\rangle \approx \frac{1}{2}\langle \hat{X}_a \hat{P}_b +   \hat{P}_b \hat{X}_a\rangle \approx 
 \frac{1}{2}\langle \hat{X}_b \hat{P}_a +   \hat{P}_a \hat{X}_b\rangle \approx  \frac{1}{2}\langle \hat{X}_b \hat{P}_b +   \hat{P}_b \hat{X}_b\rangle \approx 0,
\end{equation}
\begin{equation}\begin{split}
	\frac12 \langle \hat{X}_a \hat{X}_b + \hat{X}_b \hat{X}_a\rangle \approx & -\Gamma \Big( 2 \bar{n} (\alpha^4 \gamma_m^2 + \alpha^4 \gamma_m \Gamma + \alpha^2 \gamma_m^2 + \alpha^2 \gamma_m \Gamma) + (\alpha^4 \gamma_m^2 + \alpha^4 \gamma_m \Gamma + \alpha^2 \gamma_m^2 + \alpha^2 \gamma_m \Gamma) \\
	& \qquad + 4 (\alpha^2+1) \alpha^2 (\gamma_m +\Gamma ) \kappa _G - (\alpha^2+1) \alpha^2 \gamma_m (\gamma_m +\Gamma ) \cosh(2r) \\
	& \qquad + \sinh(2r) (\alpha^2 \gamma_m^2 + 2\alpha^2 \gamma_m \Gamma + \gamma_m^2 + 2\gamma_m \Gamma + \Gamma^2) \Big) \times D,
\end{split}\end{equation}
\begin{equation}\begin{split}
	\langle \hat{P}_a^2 \rangle \approx & \Big( 2 \bar{n} (\alpha^4 \gamma_m^3 + 2\alpha^4 \gamma_m^2 \Gamma + \alpha^4 \gamma_m \Gamma^2 - \alpha^3 \gamma_m^2 \Gamma - \alpha^3 \gamma_m \Gamma^2 + 2\alpha^2 \gamma_m^3 + 4\alpha^2 \gamma_m^2 \Gamma + 2\alpha^2 \gamma_m \Gamma^2 \\
	& \qquad - \alpha \gamma_m^2 \Gamma - \alpha \gamma_m \Gamma^2 + \gamma_m^3 + 2\gamma_m^2 \Gamma + \gamma_m \Gamma^2) + (\alpha^4 \gamma_m^3 + 2\alpha^4 \gamma_m^2 \Gamma + \alpha^4 \gamma_m \Gamma^2 \\
	& \qquad - \alpha^3 \gamma_m^2 \Gamma - \alpha^3 \gamma_m \Gamma^2 + 2\alpha^2 \gamma_m^3 + 4\alpha^2 \gamma_m^2 \Gamma + 2\alpha^2 \gamma_m \Gamma^2 - \alpha \gamma_m^2 \Gamma - \alpha \gamma_m \Gamma^2 + \gamma_m^3 + 2\gamma_m^2 \Gamma + \gamma_m \Gamma^2) \\
	& \qquad + 4 (\alpha^2+1) (\gamma_m +\Gamma ) \kappa _G (\alpha^2 \gamma_m +\alpha^2 \Gamma -\alpha \Gamma +\gamma_m +\Gamma ) \\
	& \qquad - \sinh(2r) (\alpha^3 \gamma_m^2 \Gamma + 2\alpha^3 \gamma_m \Gamma^2 + \alpha \gamma_m^2 \Gamma + 2\alpha \gamma_m \Gamma^2 + \alpha \Gamma^3) \\
	& \qquad + (\alpha^2+1) \Gamma (\gamma_m +\Gamma ) \cosh (2 r) ((\alpha^2+\alpha +1) \gamma_m +\Gamma ) \Big) \times D,
\end{split}\end{equation}
\begin{equation}\begin{split}
	\frac12 \langle \hat{P}_a \hat{P}_b + \hat{P}_b \hat{P}_a\rangle \approx & \Gamma \Big( -2 \bar{n} (\alpha^4 \gamma_m^2 + \alpha^4 \gamma_m \Gamma + \alpha^2 \gamma_m^2 + \alpha^2 \gamma_m \Gamma) - (\alpha^4 \gamma_m^2 + \alpha^4 \gamma_m \Gamma + \alpha^2 \gamma_m^2 + \alpha^2 \gamma_m \Gamma) \\
	& \qquad - 4 (\alpha^2+1) \alpha^2 (\gamma_m +\Gamma ) \kappa _G + (\alpha^2+1) \alpha^2 \gamma_m (\gamma_m +\Gamma ) \cosh(2r) \\
	& \qquad + \sinh(2r) (\alpha^2 \gamma_m^2 + 2\alpha^2 \gamma_m \Gamma + \gamma_m^2 + 2\gamma_m \Gamma + \Gamma^2) \Big) \times D,
\end{split}\end{equation}
\begin{equation}\begin{split}
	\langle \hat{X}_b^2 \rangle \approx & \Big( 2 \bar{n} (\alpha^4 \gamma_m^3 + 2\alpha^4 \gamma_m^2 \Gamma + \alpha^4 \gamma_m \Gamma^2 + \alpha^3 \gamma_m^2 \Gamma + \alpha^3 \gamma_m \Gamma^2 + 2\alpha^2 \gamma_m^3 + 4\alpha^2 \gamma_m^2 \Gamma + 2\alpha^2 \gamma_m \Gamma^2 \\
	& \qquad + \alpha \gamma_m^2 \Gamma + \alpha \gamma_m \Gamma^2 + \gamma_m^3 + 2\gamma_m^2 \Gamma + \gamma_m \Gamma^2) + (\alpha^4 \gamma_m^3 + 2\alpha^4 \gamma_m^2 \Gamma + \alpha^4 \gamma_m \Gamma^2 \\
	& \qquad + \alpha^3 \gamma_m^2 \Gamma + \alpha^3 \gamma_m \Gamma^2 + 2\alpha^2 \gamma_m^3 + 4\alpha^2 \gamma_m^2 \Gamma + 2\alpha^2 \gamma_m \Gamma^2 + \alpha \gamma_m^2 \Gamma + \alpha \gamma_m \Gamma^2 + \gamma_m^3 + 2\gamma_m^2 \Gamma + \gamma_m \Gamma^2) \\
	& \qquad + 4 (\alpha^2+1) (\gamma_m +\Gamma ) \kappa _G ((\alpha^2+1) \gamma_m + (\alpha^2+\alpha+1)\Gamma ) \\
	& \qquad - \sinh(2r) (\alpha^3 \gamma_m^2 \Gamma + 2\alpha^3 \gamma_m \Gamma^2 + \alpha \gamma_m^2 \Gamma + 2\alpha \gamma_m \Gamma^2 + \alpha \Gamma^3) \\
	& \qquad + (\alpha^2+1) \Gamma (\gamma_m +\Gamma ) \cosh (2 r) (\alpha^2 \gamma_m -\alpha \gamma_m +\gamma_m +\Gamma ) \Big) \times D,
\end{split}\end{equation}
\begin{equation}\begin{split}
	\langle \hat{P}_b^2 \rangle \approx & \Big( 2 \bar{n} (\alpha^4 \gamma_m^3 + 2\alpha^4 \gamma_m^2 \Gamma + \alpha^4 \gamma_m \Gamma^2 + \alpha^3 \gamma_m^2 \Gamma + \alpha^3 \gamma_m \Gamma^2 + 2\alpha^2 \gamma_m^3 + 4\alpha^2 \gamma_m^2 \Gamma + 2\alpha^2 \gamma_m \Gamma^2 \\
	& \qquad + \alpha \gamma_m^2 \Gamma + \alpha \gamma_m \Gamma^2 + \gamma_m^3 + 2\gamma_m^2 \Gamma + \gamma_m \Gamma^2) + (\alpha^4 \gamma_m^3 + 2\alpha^4 \gamma_m^2 \Gamma + \alpha^4 \gamma_m \Gamma^2 \\
	& \qquad + \alpha^3 \gamma_m^2 \Gamma + \alpha^3 \gamma_m \Gamma^2 + 2\alpha^2 \gamma_m^3 + 4\alpha^2 \gamma_m^2 \Gamma + 2\alpha^2 \gamma_m \Gamma^2 + \alpha \gamma_m^2 \Gamma + \alpha \gamma_m \Gamma^2 + \gamma_m^3 + 2\gamma_m^2 \Gamma + \gamma_m \Gamma^2) \\
	& \qquad + 4 (\alpha^2+1) (\gamma_m +\Gamma ) \kappa _G ((\alpha^2+1) \gamma_m + (\alpha^2+\alpha+1)\Gamma ) \\
	& \qquad + \sinh(2r) (\alpha^3 \gamma_m^2 \Gamma + 2\alpha^3 \gamma_m \Gamma^2 + \alpha \gamma_m^2 \Gamma + 2\alpha \gamma_m \Gamma^2 + \alpha \Gamma^3) \\
	& \qquad + (\alpha^2+1) \Gamma (\gamma_m +\Gamma ) \cosh (2 r) (\alpha^2 \gamma_m -\alpha \gamma_m +\gamma_m +\Gamma ) \Big) \times D
\end{split}\end{equation}
\end{subequations}
where $D = \Big[ 2 \left(\alpha ^2+1\right) (\gamma_m +\Gamma ) \left(\left(\alpha ^2+1\right) \gamma_m ^2+2 \left(\alpha ^2+1\right) \gamma_m  \Gamma +\Gamma ^2\right) \Big]^{-1}$.

The logarithmic negativity between the two oscillators $a$ and $b$ can be calculated from the covariance matrix \cite{horodecki2009quantum,woolley2014two}, as $E_\mathcal{N}=\max\{0,-\ln{2\eta}\}$, where $\eta=2^{-1/2}\{\Sigma(V)-[\Sigma(V)^2-4\det V]^{1/2} \}^{1/2}$ and $\Sigma(V)=\det V_a+\det V_b-2\det V_{ab}$. Here $V_a$, $V_b$ and $V_{ab}$ are the $2\times 2$ sub-matrices shown in Eq.~\eqref{eq:def_cov}. The result is:

\begin{equation}
\begin{split}
   &E_\mathcal{N}\approx \max \Bigg\{0,  -\ln \Bigg\{\Biggl\lbrace \frac{1}{2 \left(\alpha ^2+1\right)^2 (\gamma_m +\Gamma )^2 \left(\left(\alpha ^2+1\right) \gamma_m ^2+2 \left(\alpha ^2+1\right) \Gamma  \gamma_m +\Gamma ^2\right)^2} \\
   & \times \Biggl( \gamma_m ^6 \alpha ^8+2 \gamma_m ^3 \Gamma ^3 \alpha ^8+5 \gamma_m ^4 \Gamma ^2 \alpha ^8+4 \gamma_m ^6 \bar{n}^2 \alpha ^8+8 \gamma_m ^3 \Gamma ^3 \bar{n}^2 \alpha ^8+20 \gamma_m ^4 \Gamma ^2 \bar{n}^2 \alpha ^8+16 \gamma_m ^5 \Gamma  \bar{n}^2 \alpha ^8+4 \gamma_m ^5 \Gamma  \alpha ^8 \\
   & \qquad +4 \gamma_m ^6 \bar{n} \alpha ^8+8 \gamma_m ^3 \Gamma ^3 \bar{n} \alpha ^8+20 \gamma_m ^4 \Gamma ^2 \bar{n} \alpha ^8+16 \gamma_m ^5 \Gamma  \bar{n} \alpha ^8+4 \gamma_m ^6 \alpha ^6+\gamma_m  \Gamma ^5 \alpha ^6+\frac{17}{2} \gamma_m ^2 \Gamma ^4 \alpha ^6 \\
   & \qquad +20 \gamma_m ^3 \Gamma ^3 \alpha ^6+25 \gamma_m ^4 \Gamma ^2 \alpha ^6+16 \gamma_m ^6 \bar{n}^2 \alpha ^6+12 \gamma_m ^2 \Gamma ^4 \bar{n}^2 \alpha ^6+56 \gamma_m ^3 \Gamma ^3 \bar{n}^2 \alpha ^6+92 \gamma_m ^4 \Gamma ^2 \bar{n}^2 \alpha ^6 \\
   & \qquad +64 \gamma_m ^5 \Gamma  \bar{n}^2 \alpha ^6+16 \gamma_m ^5 \Gamma  \alpha ^6+\gamma_m  \Gamma ^5 \cosh (4 r) \alpha ^6+\frac{3}{2} \gamma_m ^2 \Gamma ^4 \cosh (4 r) \alpha ^6+2 \gamma_m ^3 \Gamma ^3 \cosh (4 r) \alpha ^6 \\
   & \qquad +\gamma_m ^4 \Gamma ^2 \cosh (4 r) \alpha ^6+16 \gamma_m ^6 \bar{n} \alpha ^6+12 \gamma_m ^2 \Gamma ^4 \bar{n} \alpha ^6+56 \gamma_m ^3 \Gamma ^3 \bar{n} \alpha ^6+92 \gamma_m ^4 \Gamma ^2 \bar{n} \alpha ^6+64 \gamma_m ^5 \Gamma  \bar{n} \alpha ^6 \\
   & \qquad +6 \gamma_m ^6 \alpha ^4+\frac{\Gamma ^6 \alpha ^4}{2}+6 \gamma_m  \Gamma ^5 \alpha ^4+20 \gamma_m ^2 \Gamma ^4 \alpha ^4+38 \gamma_m ^3 \Gamma ^3 \alpha ^4+41 \gamma_m ^4 \Gamma ^2 \alpha ^4+24 \gamma_m ^6 \bar{n}^2 \alpha ^4 \\
   & \qquad +28 \gamma_m ^2 \Gamma ^4 \bar{n}^2 \alpha ^4+104 \gamma_m ^3 \Gamma ^3 \bar{n}^2 \alpha ^4+148 \gamma_m ^4 \Gamma ^2 \bar{n}^2 \alpha ^4+96 \gamma_m ^5 \Gamma  \bar{n}^2 \alpha ^4+24 \gamma_m ^5 \Gamma  \alpha ^4 \\
   \end{split}
\end{equation}
\begin{equation}
\notag
   \begin{split}
   & \qquad +\frac{1}{2} \Gamma ^6 \cosh (4 r) \alpha ^4+2 \gamma_m  \Gamma ^5 \cosh (4 r) \alpha ^4+7 \gamma_m ^2 \Gamma ^4 \cosh (4 r) \alpha ^4+8 \gamma_m ^3 \Gamma ^3 \cosh (4 r) \alpha ^4 \\
   & \qquad +3 \gamma_m ^4 \Gamma ^2 \cosh (4 r) \alpha ^4+24 \gamma_m ^6 \bar{n} \alpha ^4+28 \gamma_m ^2 \Gamma ^4 \bar{n} \alpha ^4+104 \gamma_m ^3 \Gamma ^3 \bar{n} \alpha ^4+148 \gamma_m ^4 \Gamma ^2 \bar{n} \alpha ^4+96 \gamma_m ^5 \Gamma  \bar{n} \alpha ^4 \\
   & \qquad +4 \gamma_m ^6 \alpha ^2+\frac{3 \Gamma ^6 \alpha ^2}{2}+5 \gamma_m  \Gamma ^5 \alpha ^2+\frac{25}{2} \gamma_m ^2 \Gamma ^4 \alpha ^2+24 \gamma_m ^3 \Gamma ^3 \alpha ^2+27 \gamma_m ^4 \Gamma ^2 \alpha ^2+16 \gamma_m ^6 \bar{n}^2 \alpha ^2 \\
   & \qquad +20 \gamma_m ^2 \Gamma ^4 \bar{n}^2 \alpha ^2+72 \gamma_m ^3 \Gamma ^3 \bar{n}^2 \alpha ^2+100 \gamma_m ^4 \Gamma ^2 \bar{n}^2 \alpha ^2+64 \gamma_m ^5 \Gamma  \bar{n}^2 \alpha ^2+16 \gamma_m ^5 \Gamma  \alpha ^2 \\
   & \qquad +\frac{1}{2} \Gamma ^6 \cosh (4 r) \alpha ^2+5 \gamma_m  \Gamma ^5 \cosh (4 r) \alpha ^2+\frac{23}{2} \gamma_m ^2 \Gamma ^4 \cosh (4 r) \alpha ^2+10 \gamma_m ^3 \Gamma ^3 \cosh (4 r) \alpha ^2 \\
   & \qquad +3 \gamma_m ^4 \Gamma ^2 \cosh (4 r) \alpha ^2+16 \gamma_m ^6 \bar{n} \alpha ^2+20 \gamma_m ^2 \Gamma ^4 \bar{n} \alpha ^2+72 \gamma_m ^3 \Gamma ^3 \bar{n} \alpha ^2+100 \gamma_m ^4 \Gamma ^2 \bar{n} \alpha ^2+64 \gamma_m ^5 \Gamma  \bar{n} \alpha ^2 \\
   & \qquad +\gamma_m ^6+\gamma_m ^2 \Gamma ^4+4 \gamma_m ^3 \Gamma ^3+6 \gamma_m ^4 \Gamma ^2+4 \gamma_m ^6 \bar{n}^2+4 \gamma_m ^2 \Gamma ^4 \bar{n}^2+16 \gamma_m ^3 \Gamma ^3 \bar{n}^2+24 \gamma_m ^4 \Gamma ^2 \bar{n}^2+16 \gamma_m ^5 \Gamma  \bar{n}^2 \\
   & \qquad +16 \left(\alpha ^2+1\right)^2 (\gamma_m +\Gamma )^2 \left(\gamma_m ^2 \left(\alpha ^2+1\right)^2+2 \gamma_m  \Gamma  \left(\alpha ^2+1\right)^2+\left(3 \alpha ^2+1\right) \Gamma ^2\right) \kappa _G^2+4 \gamma_m ^5 \Gamma \\
   & \qquad +\Gamma ^6 \cosh (4 r)+4 \gamma_m  \Gamma ^5 \cosh (4 r)+6 \gamma_m ^2 \Gamma ^4 \cosh (4 r)+4 \gamma_m ^3 \Gamma ^3 \cosh (4 r)+\gamma_m ^4 \Gamma ^2 \cosh (4 r) \\
   & \qquad +4 \gamma_m ^6 \bar{n}+4 \gamma_m ^2 \Gamma ^4 \bar{n}+16 \gamma_m ^3 \Gamma ^3 \bar{n}+24 \gamma_m ^4 \Gamma ^2 \bar{n}+16 \gamma_m ^5 \Gamma  \bar{n} \\
   & \qquad +2 \left(\alpha ^2+1\right)^2 \gamma_m  \Gamma  (\gamma_m +\Gamma )^2 \left(\left(\alpha ^2+1\right)^2 \gamma_m ^2+2 \left(\alpha ^4+\alpha ^2+1\right) \Gamma  \gamma_m +\left(\alpha ^2+1\right) \Gamma ^2\right) \cosh (2 r) \left(2 \bar{n}+1\right) \\
   & \qquad +8 \left(\alpha ^2+1\right)^2 (\gamma_m +\Gamma )^2 \left(\Gamma  \left(\left(\alpha ^2+1\right)^2 \gamma_m ^2+2 \left(\alpha ^4+\alpha ^2+1\right) \Gamma  \gamma_m +\left(\alpha ^2+1\right) \Gamma ^2\right) \cosh (2 r) \right. \\
   & \qquad \qquad \left. +\gamma_m  \left(\gamma_m ^2 \left(\alpha ^2+1\right)^2+2 \gamma_m  \Gamma  \left(\alpha ^2+1\right)^2+\left(3 \alpha ^2+1\right) \Gamma ^2\right) \left(2 \bar{n}+1\right)\right) \kappa _G \Biggr) \\
   & - \frac{1}{4} \Biggl[ \frac{1}{\left(\alpha ^2+1\right)^4 (\gamma_m +\Gamma )^4 \left(\left(\alpha ^2+1\right) \gamma_m ^2+2 \left(\alpha ^2+1\right) \Gamma  \gamma_m +\Gamma ^2\right)^4} \\
   & \times \Biggl( \Biggl( 4 \times \Biggl( \gamma_m ^6 \alpha ^8+2 \gamma_m ^3 \Gamma ^3 \alpha ^8+5 \gamma_m ^4 \Gamma ^2 \alpha ^8+4 \gamma_m ^6 \bar{n}^2 \alpha ^8+8 \gamma_m ^3 \Gamma ^3 \bar{n}^2 \alpha ^8+20 \gamma_m ^4 \Gamma ^2 \bar{n}^2 \alpha ^8+16 \gamma_m ^5 \Gamma  \bar{n}^2 \alpha ^8+4 \gamma_m ^5 \Gamma  \alpha ^8 \\
   & \qquad +4 \gamma_m ^6 \bar{n} \alpha ^8+8 \gamma_m ^3 \Gamma ^3 \bar{n} \alpha ^8+20 \gamma_m ^4 \Gamma ^2 \bar{n} \alpha ^8+16 \gamma_m ^5 \Gamma  \bar{n} \alpha ^8+4 \gamma_m ^6 \alpha ^6+\gamma_m  \Gamma ^5 \alpha ^6+\frac{17}{2} \gamma_m ^2 \Gamma ^4 \alpha ^6 \\
   & \qquad +20 \gamma_m ^3 \Gamma ^3 \alpha ^6+25 \gamma_m ^4 \Gamma ^2 \alpha ^6+16 \gamma_m ^6 \bar{n}^2 \alpha ^6+12 \gamma_m ^2 \Gamma ^4 \bar{n}^2 \alpha ^6+56 \gamma_m ^3 \Gamma ^3 \bar{n}^2 \alpha ^6+92 \gamma_m ^4 \Gamma ^2 \bar{n}^2 \alpha ^6 \\
   & \qquad +64 \gamma_m ^5 \Gamma  \bar{n}^2 \alpha ^6+16 \gamma_m ^5 \Gamma  \alpha ^6+\gamma_m  \Gamma ^5 \cosh (4 r) \alpha ^6+\frac{3}{2} \gamma_m ^2 \Gamma ^4 \cosh (4 r) \alpha ^6+2 \gamma_m ^3 \Gamma ^3 \cosh (4 r) \alpha ^6 \\
   & \qquad +\gamma_m ^4 \Gamma ^2 \cosh (4 r) \alpha ^6+16 \gamma_m ^6 \bar{n} \alpha ^6+12 \gamma_m ^2 \Gamma ^4 \bar{n} \alpha ^6+56 \gamma_m ^3 \Gamma ^3 \bar{n} \alpha ^6+92 \gamma_m ^4 \Gamma ^2 \bar{n} \alpha ^6+64 \gamma_m ^5 \Gamma  \bar{n} \alpha ^6 \\
   & \qquad +6 \gamma_m ^6 \alpha ^4+\frac{\Gamma ^6 \alpha ^4}{2}+6 \gamma_m  \Gamma ^5 \alpha ^4+20 \gamma_m ^2 \Gamma ^4 \alpha ^4+38 \gamma_m ^3 \Gamma ^3 \alpha ^4+41 \gamma_m ^4 \Gamma ^2 \alpha ^4+24 \gamma_m ^6 \bar{n}^2 \alpha ^4 \\
   & \qquad +28 \gamma_m ^2 \Gamma ^4 \bar{n}^2 \alpha ^4+104 \gamma_m ^3 \Gamma ^3 \bar{n}^2 \alpha ^4+148 \gamma_m ^4 \Gamma ^2 \bar{n}^2 \alpha ^4+96 \gamma_m ^5 \Gamma  \bar{n}^2 \alpha ^4+24 \gamma_m ^5 \Gamma  \alpha ^4 \\
   & \qquad +\frac{1}{2} \Gamma ^6 \cosh (4 r) \alpha ^4+2 \gamma_m  \Gamma ^5 \cosh (4 r) \alpha ^4+7 \gamma_m ^2 \Gamma ^4 \cosh (4 r) \alpha ^4+8 \gamma_m ^3 \Gamma ^3 \cosh (4 r) \alpha ^4 \\
   & \qquad +3 \gamma_m ^4 \Gamma ^2 \cosh (4 r) \alpha ^4+24 \gamma_m ^6 \bar{n} \alpha ^4+28 \gamma_m ^2 \Gamma ^4 \bar{n} \alpha ^4+104 \gamma_m ^3 \Gamma ^3 \bar{n} \alpha ^4+148 \gamma_m ^4 \Gamma ^2 \bar{n} \alpha ^4+96 \gamma_m ^5 \Gamma  \bar{n} \alpha ^4 \\
   & \qquad +4 \gamma_m ^6 \alpha ^2+\frac{3 \Gamma ^6 \alpha ^2}{2}+5 \gamma_m  \Gamma ^5 \alpha ^2+\frac{25}{2} \gamma_m ^2 \Gamma ^4 \alpha ^2+24 \gamma_m ^3 \Gamma ^3 \alpha ^2+27 \gamma_m ^4 \Gamma ^2 \alpha ^2+16 \gamma_m ^6 \bar{n}^2 \alpha ^2 \\
   & \qquad +20 \gamma_m ^2 \Gamma ^4 \bar{n}^2 \alpha ^2+72 \gamma_m ^3 \Gamma ^3 \bar{n}^2 \alpha ^2+100 \gamma_m ^4 \Gamma ^2 \bar{n}^2 \alpha ^2+64 \gamma_m ^5 \Gamma  \bar{n}^2 \alpha ^2+16 \gamma_m ^5 \Gamma  \alpha ^2 \\
   & \qquad +\frac{1}{2} \Gamma ^6 \cosh (4 r) \alpha ^2+5 \gamma_m  \Gamma ^5 \cosh (4 r) \alpha ^2+\frac{23}{2} \gamma_m ^2 \Gamma ^4 \cosh (4 r) \alpha ^2+10 \gamma_m ^3 \Gamma ^3 \cosh (4 r) \alpha ^2 \\
   & \qquad +3 \gamma_m ^4 \Gamma ^2 \cosh (4 r) \alpha ^2+16 \gamma_m ^6 \bar{n} \alpha ^2+20 \gamma_m ^2 \Gamma ^4 \bar{n} \alpha ^2+72 \gamma_m ^3 \Gamma ^3 \bar{n} \alpha ^2+100 \gamma_m ^4 \Gamma ^2 \bar{n} \alpha ^2+64 \gamma_m ^5 \Gamma  \bar{n} \alpha ^2 \\
   & \qquad +\gamma_m ^6+\gamma_m ^2 \Gamma ^4+4 \gamma_m ^3 \Gamma ^3+6 \gamma_m ^4 \Gamma ^2+4 \gamma_m ^6 \bar{n}^2+4 \gamma_m ^2 \Gamma ^4 \bar{n}^2+16 \gamma_m ^3 \Gamma ^3 \bar{n}^2+24 \gamma_m ^4 \Gamma ^2 \bar{n}^2+16 \gamma_m ^5 \Gamma  \bar{n}^2 \\
   & \qquad +16 \left(\alpha ^2+1\right)^2 (\gamma_m +\Gamma )^2 \left(\gamma_m ^2 \left(\alpha ^2+1\right)^2+2 \gamma_m  \Gamma  \left(\alpha ^2+1\right)^2+\left(3 \alpha ^2+1\right) \Gamma ^2\right) \kappa _G^2+4 \gamma_m ^5 \Gamma \\
   & \qquad +\Gamma ^6 \cosh (4 r)+4 \gamma_m  \Gamma ^5 \cosh (4 r)+6 \gamma_m ^2 \Gamma ^4 \cosh (4 r)+4 \gamma_m ^3 \Gamma ^3 \cosh (4 r)+\gamma_m ^4 \Gamma ^2 \cosh (4 r) \\
   & \qquad +4 \gamma_m ^6 \bar{n}+4 \gamma_m ^2 \Gamma ^4 \bar{n}+16 \gamma_m ^3 \Gamma ^3 \bar{n}+24 \gamma_m ^4 \Gamma ^2 \bar{n}+16 \gamma_m ^5 \Gamma  \bar{n} \\
   \end{split}
\end{equation}
\begin{equation}
\notag
   \begin{split}
   & \qquad +2 \left(\alpha ^2+1\right)^2 \gamma_m  \Gamma  (\gamma_m +\Gamma )^2 \left(\left(\alpha ^2+1\right)^2 \gamma_m ^2+2 \left(\alpha ^4+\alpha ^2+1\right) \Gamma  \gamma_m +\left(\alpha ^2+1\right) \Gamma ^2\right) \cosh (2 r) \left(2 \bar{n}+1\right) \\
   & \qquad +8 \left(\alpha ^2+1\right)^2 (\gamma_m +\Gamma )^2 \left(\Gamma  \left(\left(\alpha ^2+1\right)^2 \gamma_m ^2+2 \left(\alpha ^4+\alpha ^2+1\right) \Gamma  \gamma_m +\left(\alpha ^2+1\right) \Gamma ^2\right) \cosh (2 r) \right. \\
   & \qquad \qquad \left. +\gamma_m  \left(\gamma_m ^2 \left(\alpha ^2+1\right)^2+2 \gamma_m  \Gamma  \left(\alpha ^2+1\right)^2+\left(3 \alpha ^2+1\right) \Gamma ^2\right) \left(2 \bar{n}+1\right)\right) \kappa _G \Biggr) \Biggr)^2 \\
   & - \Biggl( \left(\alpha ^2+1\right)^2 \left(\left(\alpha ^2+1\right) \gamma_m ^2+2 \left(\alpha ^2+1\right) \Gamma  \gamma_m +\Gamma ^2\right)^2 \\
   & \qquad \times \left(2 \gamma_m ^4 \alpha ^4+2 \gamma_m ^2 \Gamma ^2 \alpha ^4+8 \gamma_m ^4 \bar{n}^2 \alpha ^4+8 \gamma_m ^2 \Gamma ^2 \bar{n}^2 \alpha ^4+16 \gamma_m ^3 \Gamma  \bar{n}^2 \alpha ^4+4 \gamma_m ^3 \Gamma  \alpha ^4+8 \gamma_m ^4 \bar{n} \alpha ^4+8 \gamma_m ^2 \Gamma ^2 \bar{n} \alpha ^4 \right. \\
   & \qquad \qquad +16 \gamma_m ^3 \Gamma  \bar{n} \alpha ^4+4 \gamma_m ^4 \alpha ^2+\Gamma ^4 \alpha ^2+4 \gamma_m  \Gamma ^3 \alpha ^2+6 \gamma_m ^2 \Gamma ^2 \alpha ^2+16 \gamma_m ^4 \bar{n}^2 \alpha ^2+16 \gamma_m ^2 \Gamma ^2 \bar{n}^2 \alpha ^2 \\
   & \qquad \qquad +32 \gamma_m ^3 \Gamma  \bar{n}^2 \alpha ^2+8 \gamma_m ^3 \Gamma  \alpha ^2+\Gamma ^4 \cosh (4 r) \alpha ^2+16 \gamma_m ^4 \bar{n} \alpha ^2+16 \gamma_m ^2 \Gamma ^2 \bar{n} \alpha ^2+32 \gamma_m ^3 \Gamma  \bar{n} \alpha ^2 \\
   & \qquad \qquad +2 \gamma_m ^4+2 \Gamma ^4+4 \gamma_m  \Gamma ^3+4 \gamma_m ^2 \Gamma ^2+8 \gamma_m ^4 \bar{n}^2+8 \gamma_m ^2 \Gamma ^2 \bar{n}^2+16 \gamma_m ^3 \Gamma  \bar{n}^2+32 \left(\alpha ^2+1\right)^2 (\gamma_m +\Gamma )^2 \kappa _G^2 \\
   & \qquad \qquad +4 \gamma_m ^3 \Gamma +8 \gamma_m ^4 \bar{n}+8 \gamma_m ^2 \Gamma ^2 \bar{n}+16 \gamma_m ^3 \Gamma  \bar{n}+4 \left(\alpha ^2+1\right)^2 \gamma_m  \Gamma  (\gamma_m +\Gamma )^2 \cosh (2 r) \left(2 \bar{n}+1\right) \\
   & \qquad \qquad \left. +16 \left(\alpha ^2+1\right)^2 (\gamma_m +\Gamma )^2 \left(2 \bar{n} \gamma_m +\gamma_m +\Gamma  \cosh (2 r)\right) \kappa _G\right) \Biggr)^2 \Biggr) \Biggr]^{1/2} \Biggr\rbrace^{1/2}
   \Bigg\}.
   \end{split}
\end{equation}
The results were calculated and generated using programs. Furthermore, by performing a second-order Taylor expansion in $\alpha$, we obtain the expression for the logarithmic negativity given in Eq.~(5) of the main text.

\section{Parameters}
In Table~\ref{tab:para}, we summarize all the parameters that we have used in our model, including their definitions and reference values. Some parameter values that do not affect the simulation results are omitted. Parameter values marked with * are varied in the simulations for Fig. 2 of the main text, and values marked with ** are varied in the simulations for Fig. 3. For these adjustable parameters, only a single default value is listed. For example, the value 0 of $K_{\text{others}}$ is marked by **, means that for Fig.~2 of the main text, it is set to be 0, and  for Fig.~3 of the main text, it is varied. Some parameters are, in principle, also varied; however, their variation is so small that it has no effect on the simulation results, so we treat them as fixed—e.g., $\omega_{a(b)}'$.

\begin{table}[b]
\label{tab:para}
\centering
\renewcommand{\arraystretch}{1.8}
\caption{Parameter table}
\begin{ruledtabular}
\begin{tabular}{p{2cm} p{11cm} p{3cm}} 
\hline
Parameter & Definition & Reference Value \\
\hline
$r$ & Radius of the spherical oscillators & $0.25~\si{mm}$ \\
$d$ & The equilibrium distance between the centers of mass of the two spherical oscillators & $0.5~\si{mm}$ \\
$M$ & Mass of the spherical oscillators & $1.3~\si{mg}$ \\
$\rho$ & The density of the spherical oscillators & $19.3~\si{g/cm}^3$ \\
$\Lambda$ & The geometry-dependent form factor of the oscillators & $\pi/3$ \\
$\omega_{a(b)}$ & Mechanical frequency & $2\pi\times (50\pm0.5)$~\si{Hz} \\
$\gamma_{a(b)},\gamma_m$ & Mechanical dissipation & $^*2\pi\times2\times 10^{-9}~\si{Hz}$ \\
$K_G$ & The gradient of the gravitational force between the two oscillators & $1.7\times 10^{-12}~\si{N/m}$ \\
$K_\text{others}$& The gradient of the non-gravitational forces between the two oscillators. & $^{**}0$ \\
$K_M$ & The gradient of the total forces between the two oscillators, $K_M=K_G+K_\text{others}$ & $^{**}1.7\times 10^{-12}~\si{N/m}$ \\
$\omega_{a(b)}'$ & The corrected frequency of the mechanical oscillator, $\omega'_{a (b)} = \sqrt{\omega_{a (b)}^2 - K_M / M}$  & $2\pi\times (50\pm0.5)$~\si{Hz}  \\
\hline
\end{tabular}
\end{ruledtabular}
\end{table}

\begin{table}[t]
\centering
\renewcommand{\arraystretch}{1.8}
\begin{ruledtabular}
\begin{tabular}{p{2cm} p{11cm} p{3cm}} 
\hline
Parameter & Definition & Reference Value \\
\hline
$x_\text{zpf,a(b)}$ &  The zero point fluctuations of the oscillators, $x_{\text{zpf},a(b)} = \sqrt{\hbar / 2 M {\omega'}_{a(b)}}$ & $3.6\times 10^{-16}$~\si{m}\\
$k_G$ & $k_G = K_G x_{\text{zpf},a}x_{\text{zpf},b} /\hbar$ & $2\pi\times 7\times 10^{-10}~\si{Hz}$  \\
$k_\text{others}$ & $k_\text{others} = K_\text{others} x_{\text{zpf},a}x_{\text{zpf},b} /\hbar$  & $^{**}0$  \\
$k_M$ & $k_M = K_M x_{\text{zpf},a}x_{\text{zpf},b} /\hbar$  & $^{**}2\pi\times 7\times 10^{-10}~\si{Hz}$  \\
$\omega_c$ & The frequency of the microwave cavity mode & - \\
$\kappa$  & The dissipation of the microwave cavity mode & $2\pi\times 10~\si{Hz}$ \\ 
$g_{a(b)},g$ & The single-photon optomechanical coupling strengths & $2\pi\times 1~\si{Hz}$ \\
$\omega_m$ & The central frequency of the mechanical oscillators, $\omega_m=(\omega_a'+\omega_b')/2$  & $2\pi\times 50~\si{Hz}$ \\
$Q_m$ & The mechanical quality factor, $Q_m=\omega_m/\gamma_m$ & $^*2.5\times 10^{10}$ \\
$\mathcal{E}_{\pm}$ & The pump amplitudes & $^*|\mathcal{E}_+|=100~\si{Hz},\ \ \ \ \ \ \ \ \ \ $ $|\mathcal{E}_-|=200~\si{Hz}$  \\
$\overline{n}_{a(b)},\overline{n}$ & The thermal phonon occupation number  & $4\times 10^6$ \\
$\Gamma_{a(b)}$ & The measurement rate in the KTM model& - \\
$\kappa_{G,a(b)},\kappa_G$ &  The additional dissipation rates introduced by the classical gravity model, $\kappa_{G,a(b)} = [\Gamma_{a(b)} /\hbar + K_G^2/4\hbar\Gamma_{b(a)}]x_{\text{zpf},a(b)}^2$& $2\pi\times 7\times 10^{-10}~\si{Hz}$ \\
$\Omega$ & The frequency difference between the two oscillators, $\Omega = (\omega'_a - \omega'_b)/2$ & $2\pi\times 0.5~\si{Hz}$ \\
$\overline{c}_{\pm}$ & The steady-state
amplitudes of the fields at the driven sidebands, $\overline{c}_\pm = i\mathcal{E}_\pm / (\pm i\omega_m - \kappa/2)$ & $^*\overline{c}_+\approx 0.3,\overline{c}_-\approx 0.6$ \\
$G_{\pm}$ &  The many-photon optomechanical couplings, $G_\pm = (g_a + g_b)\overline{c}_\pm /2$ & $^*G_+\approx 2.0~\si{Hz},G_-\approx 4.0~\si{Hz}$ \\
$r$ & $r = \tanh^{-1} (G_+ / G_-)$ & $0.55$ \\
$\mathcal{G}$ & The effective optomechanical coupling, $\mathcal{G} = \sqrt{G_-^2 - G_+^2}$  & $^*3.4~\si{Hz}$ \\
$\Gamma$ & The optomechanical damping rate, $\Gamma = 4\mathcal{G}^2 / \kappa$ & $^*0.76~\si{Hz}$ \\
$T$ & The environmental temperature & $10~\si{mK}$ \\

\hline
\end{tabular}
\end{ruledtabular}
\end{table}

\section{Approximate proportionality between entanglement difference and quality factor
}

In order to obtain the result shown in Eq.~(6) of the main text, where the entanglement difference between the quantum and classical models is approximately proportional to the quality factor $Q_m$, we further approximate Eq.~(5) in the main text. First, we keep only the first term inside the logarithm (i.e., assuming $k_M/\Omega$ is negligible), and then perform a first-order Taylor expansion of $\kappa_G$, yielding:

\begin{equation}
    E_{\mathcal{N}}\approx-\ln\bigg[\frac{(2\overline{n}+1)\gamma_m +4\kappa _G+\Gamma  e^{-2 r}}{\gamma_m +\Gamma }\bigg]\approx -\ln\bigg[\frac{(2\overline{n}+1)\gamma_m +\Gamma  e^{-2 r}}{\gamma_m +\Gamma }\bigg] 
- \frac{4\kappa_G}{(2\overline{n}+1)\gamma_m +\Gamma  e^{-2 r}}.
\end{equation}

By subtracting the result of the classical model (\(\kappa_G\neq0\)) from that of the quantum model (\(\kappa_G= 0\)), we obtain:

\begin{equation}
    E_{\mathcal{N},q}-E_{\mathcal{N},c}=E_\mathcal{N}|_{\kappa_G=0}-E_\mathcal{N}\approx \frac{4\kappa_G}{(2\overline{n}+1)\gamma_m + \Gamma e^{-2r}} \propto \frac{1}{\gamma_m} \propto Q_m.
\end{equation}
The first proportionality sign follows from our parameter-tuning rule: keeping the ratio \(\gamma_m/\Gamma\) fixed, so that \(\Gamma\propto\gamma_m\). The second one uses \( Q_m = \omega_m/\gamma_m \) together with the fact that we keep \( \omega_m \) fixed.

\bibliography{apssamp}